\documentclass{article}
\usepackage{graphicx,bm,color,caption,subcaption,amsthm,cleveref,epstopdf}
\usepackage[]{hyperref}
\usepackage{cite}
\hypersetup{colorlinks,citecolor=blue,filecolor=black,linkcolor=blue, urlcolor=black}
\usepackage{soul}
\usepackage{etoolbox}

\newcommand {\etal}{{\it et al.$\,$}}

\renewcommand{\epsilon}{\varepsilon}

\newcommand{\rw}{\rightarrow}
\newcommand{\bea}{\begin{eqnarray}}
\newcommand{\eea}{\end{eqnarray}}

\newcommand{\bfr}{{\bf r}}

\newcommand{\bfb}{{\bf b}}
\newcommand{\gav}{\langle\gamma(t)\rangle}
\newcommand{\dg}{\dot\gamma}

\graphicspath{{./}}

\title{Avalanches and Plastic Flow in Crystal Plasticity: An Overview}
\author{Stefanos Papanikolaou$^{1,2,3}$, Yinan Cui$^4$ and Nasr Ghoniem$^4$}
\date{
	$^1$Department of Mechanical and Aerospace Engineering, Western Virginia University, West Virginia 26506\\
	$^2$Department of Physics, Western Virginia University, West Virginia 26506\\
	$^3$Department of Mechanical Engineering, Johns Hopkins University, Baltimore, MD, 21218\\
	$^4$Department of Mechanical and Aerospace Engineering, University of California, Los Angeles, CA 90095\\
	\today
	}

\begin{document}
\maketitle
\newpage
\tableofcontents
\newpage

\begin{abstract}
Crystal plasticity is mediated through dislocations, which form knotted configurations in a complex energy landscape.  Once they disentangle and move, they may also be impeded by permanent obstacles with finite energy barriers or frustrating long-range interactions. The outcome of such complexity is the emergence of dislocation avalanches as the basic mechanism of plastic flow in solids at the nanoscale. While the deformation behavior of bulk materials appears smooth, a predictive model should clearly be based upon the character of these dislocation avalanches and their associated strain bursts.  We provide here a comprehensive overview of experimental observations, theoretical models and computational approaches that have been developed to unravel the multiple aspects of dislocation avalanche physics and the phenomena leading to strain bursts in crystal plasticity.
\end{abstract}
\textbf{keywords}\\
dislocations, \quad avalanches, \quad strain bursts, \quad cracks, \quad surface steps.

\newpage
\pagenumbering{arabic}
\setcounter{page}{1}
\pagestyle{plain}
\section{Introduction}
A great early success of the statistical mechanics of systems in thermodynamic equilibrium was the establishment of clear connections between macroscopic fluid flow and the microscopic motion of its molecules. In the case of equilibrium liquids, there exists a link between the Brownian movement of microscopic molecules and the macroscopic flow of the liquid, that is summarized in Einstein's relation~\cite{Landau:1958qy}. Moreover, the theory of gases and liquids was enhanced when the law of corresponding states was reported in 1873 by van der Waals: This empirical law states that at high enough pressure and temperatures, the equation of state for dozens of real gases contain {\it only} $3$ fitting parameters, which can be expressed through {\it reduced} or {\it scaling} variables for temperature $T/T_c$, pressure $P/P_c$ and volume $V/V_c$. Back in the early 1900s, the law of corresponding states was known to hold close to the critical point $(T_c,P_c,V_c)$ when in the gas phase, but did not hold in the liquid state, and also not for extremely high temperatures. However, what remains as most remarkable is the fact that, 100 years after its discovery, in 1972, the theory of the renormalization group  was developed to generalize and explain the law of corresponding states in gases in a systematic way, from first principles and without using {\it ad hoc} assumptions \cite{wilson1972critical}. This theory presents a general and systematic way of coarse-graining continuum theories in an effort to unify behaviors across systems with various interatomic potentials. In a general sense, renormalization group can be formally used near a continuous phase transition, describe local observables in scaling forms, that involve a reduced number of variables, and  predict the behavior of all nearby phases through the extensive study of just one of them. The existence of such multiscale modeling near continuous phase transitions has allowed for remarkable progress in a wide range of topics across science and engineering. In crystal plasticity, such a theory would allow, for example, to understand Stage I crystal plasticity only by just the careful study of the elastic behavior.

Our understanding of plastic deformation in solids has emerged as an extension of fluid behavior, where the term  ``plastic flow" has been coined mainly through the work of Orowan to describe plasticity as some kind of fluid motion.  However, it still remains a remarkable hurdle to develop a general, systematic understanding of crystal plasticity that would predict stress-strain responses for many metals at a range of temperatures and pressures using ``scaling" or ``reduced " variables. The possibility of generalizing the understanding of renormalization group in liquids to include crystal plasticity, has been driving the effort in this review. In what follows, we aim at describing the range of phenomena and continuum theories that point towards developing a systematic multiscale understanding of crystal plasticity.  

It is an interesting exercise to associate crystal plasticity with dynamical critical behavior, such as spinodal decomposition.  If one distributes various dislocation components (say, positive and negative edge in single slip) in a random, homogeneous mixture, as for example in the model of Ref. \cite{groma2003spatial}, then it is plausible to imagine that the ``demixing" process of these components under mechanical stress might resemble spinodal decomposition. However, this interesting academic analogy appears too simple to explain the reality of crystal plasticity. The fundamental components of plasticity are extended dislocation defects, for which the corresponding ensembles emerge from a variety of possible initial patterns, and also, dislocations multiply in various ways, and form junctions as they propagate.

Naturally, one can consider dislocations as non-interacting, and using this starting point Orowan proposed that the basic dynamics of dislocations are akin to that of fluids, namely local strain rates are given, for any position $\bfr$, by $\dot\gamma(\bfr)=\rho(\bfr,t) b v(\bfr,t)$, where $v$ is the local dislocation velocity and $t$ is time. However, dislocations mutually interact both at short distances (through the formation of junctions) as well as at long distances through long-range elastic interactions, and their motion is approximately athermal at temperatures away from melting. As a consequence, dislocations do not perform Brownian walk at the nanoscale, and they remain at a state far from thermodynamic equilibrium, even at relatively high temperatures. While Orowan's hypothesis has always been somewhat unproven, it has been used readily by all kinds of coarse-graining continuum theories up to the current day. However, it has become clear over the last decade, through extensive simulations and experiments, that it dramatically fails at the nanoscale.~\footnote{The failure of Orowan's hypothesis does not relate to the obvious identity of $\dot\gamma(\bfr)$ with the sum of individual dislocation slip rates. Instead,  the failure is in the simple coarse-grained weighting of $\rho(\bfr,t)$ as it happens in typical fluids. A consequence of this failure amounts to the  various suggestions that ``$\rho(\bfr,t)$" corresponds to distinct dislocation sets, such as the statistically stored or geometrically necessary ones.~\cite{Rice:1970fj,Roy:2005yq}.} \emph{Dislocations at the nanoscale do not move smoothly, but instead they transition through bursts of activity}.

Orowan hypothesis' failure through the observation of non-smooth plastic deformation at the nanoscale has led to various generalizations of dislocation dynamics that consequently led to novel predictions for microscale crystal plasticity. The most natural way to model such behavior is through using the analogy of a rubber band traversing, under an applied force, through a landscape of randomly placed pins. This picture may resemble the behavior of non-interacting long glide dislocations as they attempt to cross a dislocation forest \cite{Mohles:1999zr}. Thus, generalizing Orowan's assumption, dislocations may be independent and non-interacting but they may also be kinetically ``arrested" through junctions that are distributed along the glide direction.  In the simplest version of this elastic interface depinning picture, as we will review in Section~\ref{sec:model}, the global yield stress should depend on the disorder distribution, defined through the variance of the dislocation back-stress $\langle\delta\tau_b\rangle^2$ as (in the infinite-range approximation) $\tau_c\sim G \langle \delta \tau_b \rangle^2 / \langle \tau_b\rangle$, where $G$ is the shear modulus and $\tau_b$  the stochastic material/dislocation back stress. In this case, dislocation flow would satisfy the depinning law \cite{fisher1985sliding,Nattermann:1992vn} for $\tau>\tau_c$: $
\dot\gamma\sim(\tau - \tau_c)^{\beta}$,  where $\tau_c$ is the critical depinning shear stress, and $\tau$ is applied shear stress, while $\beta$ is an exponent that is fully controlled by the dynamics of dislocations. If the jerkiness of collective dislocation motion inside a single avalanche occurs at length scales up to $\xi\sim1/(\tau-\tau_c)^\nu$ and timescales $\delta t_\xi \sim \xi^z$ while the dislocation motion is smoother on longer scales, then the overall rate of dislocation motion is $\xi^{\zeta}/\delta t_\xi$. Thus, $\beta=(z-\zeta)\nu$. For reference, $\beta=1,\zeta=0,z=1,\nu=1$ for infinite-range elastic isotropic interactions between dislocations.

While this result appears phenomenologically similar to typical engineering crystal plasticity laws, its origin is radically different and the main parameter $\beta$ is set by basic principles, instead of exhaustive material behavior investigation.  A defining difference between the depinning description and Orowan's one is that the actual dynamics required for the latter is assumed to be smooth while the former abrupt, as it is experimentally observed \cite{greer2011plasticity,uchic2009plasticity}. Clearly, the power of a microscopically consistent continuum theory would be the predictability of non-trivial, interaction-driven exponents, such as the rate-sensitivity parameter in continuum plasticity modeling $m\equiv 1/\beta$. The typical continuum plasticity formulation is $\dot\gamma_\alpha=\dot\gamma_0 \frac{\tau_{(\alpha)} - \tau_b}{g_\alpha} \left|  \frac{\tau_{(\alpha)} - \tau_b}{g_\alpha} \right|^{1/m-1} $, where $m$ is the strain rate sensitivity parameter, $\dot\gamma_0$ is a reference strain rate, and $g_\alpha(\gamma)$ is the appropriate hardness parameter that depends on the accumulated shear strain $\gamma$. Rate-independent behavior arises for $m\rw0$~\cite{Asaro:1985rt}. Assuming positive loads, it becomes clear that $\dot\gamma_\alpha\sim\frac{1}{g_\alpha} (\tau_{(\alpha)} - \tau_b)^{1/m}$, thus one could assimilate $\beta\equiv 1/m$.

While the generalization to independent dislocations that traverse through a strongly disordered environment is natural, it is clearly a far cry from the reality of plastically deforming crystals. Gliding dislocations almost always have a loop topology and they may multiply when pinned; an example arises through the Frank-Read source nucleation mechanism~\cite{Asaro:2006fr}. Furthermore, dislocations have a finite length between junctions that varies strongly as function of the average dislocation density.  Due to this strong variety in dislocation lengths, it is possible that dislocations that are relatively (to the sample size) long (like the aforementioned glide depinning ones) and therefore, ``near critical" always emerge in pristine or small crystals. This possibility has raised the possibility of the realization of self-organized criticality (SOC) in small crystals.~\cite{uchic2009plasticity,bak1987self}. Also, it is possible to use the critical elastic interface depinning analogy in a loose, conceptual manner and conjecture that its basic predictions may hold true. One may conceptually think of the dislocation ensemble as a whole in terms of an elastically coupled medium that moves along the deformation direction, while pinned on strong disorder (e.g. junctions and obstacles). However,  the extreme complexity of dislocation ensembles requires a variety of further generalizations before theoretically justifying the SOC or elastic depinning picture; such generalizations  have led to a plethora of models that have addressed particular features of crystal plasticity. 

The necessity to model junction/obstacle formation/annihilation, a critical aspect of dislocation dynamics, requires further generalizations of modeling approaches: A possible way is through the assumption that a set of point dislocations may move through phenomenological dynamics that is driven by junction/obstacle formation/annihilation and then, one is led to reaction-diffusion models \cite{walgraef1985dislocation,aifantis1984microstructural}. Then, one may even proceed further to include dynamical terms (that may originate to junction/obstacle aging or other relaxation processes) that allow for highly non-linear dynamical phenomena, such as relaxation oscillators and limit cycles~\cite{desroches2012mixed}. Such models have been studied~\cite{ananthakrishna2007current} and have explained a variety of behaviors that arise in the context of dynamic strain aging and the Portevin - Le Chatelier effect \cite{cottrell1949dislocation,cottrell1953lxxxvi}. Dynamical phenomena may also be included in the context of elastic interface depinning models, leading to the identification of relaxation oscillations and stick slip phenomena~\cite{papanikolaou2012quasi,Braun:2002sf,Jagla:2007vf}.

Finally, the most direct way of treating the dislocation ensemble complexity is through the numerical simulation of the evolution of an initial configuration of dislocations that arises from the application of external stress, and continuously minimizing the crystalline elastic energy \cite{kubin2013dislocations}. This direct simulation method (Discrete Dislocation Dynamics (DDD)) \cite{Weygand:2002mz} offers several advantages, since the energetics of junctions, obstacles and dislocations are captured to a rather high resolution. However, the computational complexity limits the applicability of this method to small volumes. Since the fluid character of dislocation ``flow" is challenged at the microscopic level, there is a necessity of statistical mechanics methods that address issues of sampling the initial microstructure in  self-consistent ways. This problem resembles analogous  problems where either disorder is intrinsically large ({\it e.g.} amorphous solids or granular piles under shear)~ \cite{Makse:2002gf,Papanikolaou:2013ul} or the relative volume is small (as for the gas flow where the Knudsen number is larger than 0.1).\cite{Cercignani:2000pd}

In the following, we will explore recent progress in unraveling the dynamics of {\it mutually dependent} dislocation ensembles.  In Section~\ref{sec:exp}, we will review experimental evidence of abrupt behavior at the nano- and micro-scales through the findings in: i) uniaxial compression of micropillars, ii) acoustic energy emission during tension/compression of macroscopic single and poly-crystals, iii) pop-in events and noise during nanoindentation, iv) serrations in the Portevin-Le Chatelier effect, v) the depinning effect known as the  L\"uders band phenomenon, strain bursts in vi) irradiated materials, vii) in precipitation-hardened materials, viii) and during cyclic loading, and finally,  ix) the observation of plausible connections between strain bursts and patterning effects along the sample/pillar surface. In Section~\ref{sec:model}, we will seek the basic theoretical assumptions and models for the dynamics of dislocation ensembles, that aim to generalize the Orowan hypothesis towards explaining the non-smooth aspects of crystalline, plastically deforming behavior. We will start with the assumption of a single dislocation in a disordered landscape and inspect the applicability of this model to crystal plasticity. We will then discuss reaction-diffusion models and distribution function approaches for the modeling of abrupt events, and finally examine the assumptions and basic modeling features in 3D-DDD simulations. In Section~\ref{sec:results}, we will discuss various  aspects of the applicability of different models.
Finally, in Section~\ref{sec:pf} , we will discuss the possible connection of nanoscale dislocation dynamics with precursor signatures of incipient material failure and fracture. We will discuss various observations and probe possible future steps that may lead to the design of failure-resistant materials. We will finally conclude by discussing theoretical challenges that currently require further attention, in order to improve the insight into the abrupt dynamical behavior of dislocations at the nano- and micro-scales.

\section{Experimental Observations of Abrupt Plastic Events}
% Describe all event-like observations
\label{sec:exp}
	\subsection{Micropillars}
	With the rapid development of experimental technology for material testing at the microscale (e.g.~focused ion beam fabrication and in-situ transmission electron microscopy (TEM)), a deeper understaning of plastic deformation has emerged. Significant intermittent plastic behavior is observed during uniaxial loading of  micro- and nano- pillars \cite{uchic2004sample,greer2005size,dimiduk2006scale}. The stress-strain relationship is characterized by discrete strain steps under stress control conditions, and by intermittent serrated stress drops under strain control. The general observation is that the magnitude of the strain bursts $S$ is found to follow a power-law distribution $P(S)\sim S^{\delta}$: The power law exponent $\delta$ is reported as -1.6 in Ni pillars with characteristic length $18 \sim 30.7~\mu m$  \cite{dimiduk2006scale},  $-1.47 \sim -1.67$ in $0.8 \sim 6.3~ \mu m$ diameter Al pillars  \cite{ng2008stochastic},  -1.5 in $155 \sim 1000$ nm diameter Au pillars, and  $-1.5 \sim -1.55$ for 150 nm $\sim$ 5.82 $\mu$m diameter Mo pillars \cite{brinckmann2008fundamental,zaiser2008strain}. For 1 $\sim 20~ \mu m$ LiF crystals, a larger power law exponent of $1.8 \sim 2.9$ is found \cite{dimiduk2010experimental}.  A detailed summary of the recorded burst quantity and the power law exponent in micro/nano pillars is given in Table 3 of reference \cite{cui2016Influence}, and in \cite{kubin2013dislocations}. 
	
 It is generally believed that dislocation avalanches result from the collective and correlated dynamics of dislocations \cite{csikor2007dislocation,wang2012sample}. By monitoring the displacement jump velocity during displacement-rate-controlled micro compression testing, the internal stress field landscape is suggested to control the collective motion of dislocations \cite{maass2013small}. On the other hand, in sub micron crystals, the characteristic lengths of dislocation sources are short due to the truncation effect of small pillars. At the same time, the dislocation source density is also very limited. TEM observations provided direct evidence into source-controlled dislocation plasticity \cite{oh2009situ,kiener2011source}. The operation and inactivation of dislocation sources is sometimes found to directly correlate with the occurrence of strain bursts \cite{rao2008athermal,tang2008dislocation,cui2014theoretical}. Based on time-resolved Laue diffraction, strain bursts were found to be also associated with the strain gradient in the internal local region, as well as with complex dislocation interactions \cite{maass2007time}. In addition, experiments and simulations inferred that intermittent plasticity was related to the easy dislocation annihilation from free surfaces in small-scale crystals. Introducing a barrier to surface annihilation was found to inhibit strain bursts, such as by coating the surface \cite{ng2009effects,el2011trapping,cui2015theoretical,li2017small}. Similar inhibition of strain bursts was also observed by introducing grain boundaries \cite{richeton2005breakdown,zhang2013strain,niiyama2016barrier}.

\subsection{Acoustic Emission}
During plastic deformation, energy is released through dissipation, heat generation, and traveling acoustic waves \cite{zaiser2006scale}. If the released energy is large enough, audible sounds are produced, similar to those associated with cracking \cite{swindlehurst1973acoustic}. Thus, acoustic emission signals may arise from dislocation or crack motion; here, we focus on dislocations. The released energy travels through high-frequency stress waves, which are received by sensors and then converted into a voltage. This voltage is electrically amplified and further processed as an acoustic emission signal.

To quantitatively interpret acoustic emission signals, researchers proposed various models to relate plastic deformation to dislocation motion \cite{fisher1967microplasticity,scruby1981origin,rouby1983acoustic, richeton2005dislocation}. The widely accepted notion is that the acoustic wave amplitude $A$ is proportional to the plastic strain rate \cite{weiss2007evidence}, while the acoustic energy $E_A$, defined as $E_A=\int A^2(t)dt$, scales as $E_A \sim A^2$ \cite{weiss2006seismology}. $E_A$ is thought to be proportional to the dissipated energy during dislocation avalanches \cite{weiss2000statistical}. When dislocations move uniformly, the produced acoustic emission signal is uniform and continuous, without an acoustic burst. In contrast, for spatially and temporally heterogeneous dislocation motion, acoustic bursts are observed \cite{gillis1972dislocation}. 

Power-law distributed intermittent plasticity was initially recognized in the acoustic emission experiments on single- and polycrystal ice by Weiss {\it et al.}\cite{weiss1997acoustic,weiss2000statistical, miguel2001intermittent,weiss2007evidence}. The uniqueness of ice is that its transparency makes it easy to exclude the acoustic emission activity induced by microcraking. Meanwhile, the plasticity of ice is dominated by dislocation glide from relatively low temperatures to almost the melting point. Under uniaxial compression creep conditions, the acoustic burst in single crystal ice is found to exhibit a power-law probability distribution with an exponent -1.6 for $E_A$ and -2 for $A$ \cite{miguel2001intermittent}. Since grain-boundary sliding was demonstrated not to contribute to acoustic emission \cite{frydman1975acoustic,mintzer1978acoustic}, dislocation avalanche behavior in polycrystals can also be revealed by acoustic emission testing.  In polycrystalline ice samples \cite{richeton2005breakdown}, grain boundaries were found to inhibit dislocation avalanche propagation and temporally push the system into a supercritical state. 

Apart from ice, acoustic emission experiments were also carried out on metallic crystals. The acoustic emission amplitude distribution was found to follow a power law with exponent about -2.0 in hexagonal cadmium and zinc single crystals (power law exponent is about -1.5 for $E_A$) \cite{richeton2006critical}, as well as FCC single crystal Cu \cite{weiss2007evidence}, even when multi-slip and forest hardening occur. Recently, an acoustic emission transducer was mounted on the compressed Al-5\% Mg micropillars. The acoustic signal revealed the correlation between the cooperative motion of dislocations, the stress drops under pure strain-controlled tests, and the acoustic emission signals \cite{hegyi2016micron}.

	\subsection{Nano-indentation }
Nano-indentation is a useful experimental tool to probe dislocation activity in a precise and local manner. If the region surrounding the indenter is dislocation-free, then nucleation of fresh dislocations would be expected to lead to abrupt changes in indenter displacement.  Such rapid drops have been labeled as ``pop-in" events. It has been established that in pristine/annealed FCC or BCC crystals, a large strain burst takes place in nano-indentation experiments in a somewhat reproducible applied load, of relatively predictable character. These ``pop-in" events should be distinguished from the avalanche bursts we discussed above. For small indenters, the pop-in occurs at high stress, close to the theoretical limit ($\approx G/2\pi$) \cite{courtney2005mechanical}). This is explained as a result of the limited primary indentation zone, which makes it difficult to activate dislocation sources \cite{syed1997nanoindentation,pathak2015spherical}; or by being dislocation-free, requiring dislocation nucleation at the theoretical strength \cite{morris2011size}. For initially dislocation-free materials, some experiments inferred that the first ``pop-in" is induced by collective dislocation nucleation \cite{syed1997nanoindentation,gouldstone2000discrete,shibutani2007nanoplastic}. However, Minor {\it et al.} observed that dislocations nucleated and glided at very small force, and that the pop-in was induced by the rearrangement of dislocation configurations and/or activation of dislocations on less favorable slip system  \cite{minor2006new}.  By contrast, in crystals that are pre-strained or cold-worked, it is very common to observe ``noise" during nano-indentation, which appears to be intrinsic to the material behavior. In addition, the pop-in stress is size dependent, because larger indenters correspond to a more extensive region of high stress, and a higher probability of finding pre-existing mobile dislocations \cite{morris2011size}.

	\subsection{Portevin-LeChatelier Effect} 
	Experiments consistently show that strain bursts and dislocation avalanches are prevalent features of plastic deformation at the nano- and micro-scales. Weiss {\it et al.} proposed that  dislocation avalanches spread over lamellar structures with a fractal dimension $D$ ($D<3$), so the cutoff strain increment scales as $d^{D-3}$, which decreases with increasing sample size. This explains why it is difficult to observe discontinuous and intermittent deformation in bulk samples \cite{weiss2007evidence}.  However, in solute hardening alloys, if the diffusion velocity of solute atoms is comparable to dislocation motion velocity, significant intermittent plasticity is also observed. The intermittent plastic flow behavior in large samples is a result of dislocations being repeatedly locked and unlocked from diffusing solute atoms through dynamical strain aging  \cite{kubin2002collective}. This phenomenon is known as the Portevin-Le Chatelier (PLC) effect, where correlated slip bands in a strain burst appear randomly somewhere along the crystal \cite{neuhauser1993observation}. The classical PLC effect leads to plastic
oscillations, which may be found in metallic alloys,
solid solutions, and intermetallic compounds deformed in certain
ranges of temperatures, stresses and
 strain rates
 \cite{portevin,plc-24,plc-95,plc-96}.
PLC effect is also referred to as repeated yielding, serrated or
jerky flow. This effect involves localization
of plastic deformation and propagation
of deformation bands, which lead to complex
spatio-temporal behavior.

	Statistical analysis of the stress drop magnitude shows that it exhibits power law distribution with an exponent -1.1 in single crystals of Cu$-10\%$Al  \cite{ananthakrishna1999crossover}, and about $-1 \sim -1.5$ in single- and poly crystal Al-Mg alloys (the exponent decreases with increasing solute density) \cite{lebyodkin2000spatio}. Note that the available dynamical range is significantly reduced compared with the acoustic emission of ice, suggesting the inhibition role of solutes on strain bursts, similar to the effect of grain boundaries \cite{weiss2006seismology}.  Effectively, the stress
response of the deformed specimens occurs in the form of periodic
strain bursts, and the corresponding temporal pattern involves two
different  time-scales, typical of relaxation oscillations:
a short burst period, and a longer time interval between
bursts.

\subsection{The L\"uders Phenomenon}
 The occurrence of the PLC effect requires a certain diffusion mobility of solute atoms so that they can accumulate in the dislocation stress field, and thus takes place at relatively elevated temperatures in substitutional alloys.  A similar phenomenon of smooth propagation of deformation bands was observed in low-carbon steels and certain Al-Mg alloys by L\"uders \cite{mason1910luders}. However, these ``L\"uders'' bands propagate only once in the specimen, while PLC bands can propagate repeatedly from one end of the sample to the other. Before the initiation of the L\"uders band, the nominal stress-strain curve develops a yield drop, and while the band is propagating, the corresponding undeformed material moves with constant cross head velocity, and the nominal stress-strain curve is flat.
 
 After the upper yield point is reached, collective depinning and significant dislocation motion occurs, producing strain softening. Afterwards, the plastic front advances by addition of new slip bands parallel to the old ones at the lower yield stress (also termed as "propagation stress"), until the sample is uniformly deformed at a certain strain (L\"uders strain $\varepsilon_L$). The L\"uders phenomenon was initially observed in mild steels and BCC polycrystals  \cite{luders1860dinglers}, and then widely observed in Cu-Zn, Cu-Al and other alloy crystals  \cite{neuhauser1993observation}. While PLC bands  are always initiated at the same grip of a tensile specimen, L\"uders bands do not show preference for the starting point. Another distinguishing feature is that the slope of the stress-strain curve during propagation of a L\"uders band is essentially zero, while it is positive during PLC band propagation.
% \begin{figure}[htbp]
%\centering \includegraphics[width=.6\textwidth]{Luders_replot.pdf}
%\caption{Schematic showing the yield drop and the corresponding L\"uders band propogation\cite{ananthakrishna2007current}} \label{Luders}
%\end{figure}

	\subsection{Irradiated Materials}
	After materials are subjected to irradiation, numerous defect clusters are produced. They serve as  barriers to dislocation motion. However, compared with strong indestructible precipitate bariers,  complex interactions between dislocations and irradiation defects further complicate the picture. Taking interstitial loops as an example, they may be swept by  gliding dislocations when the Burgers vector of the loop belongs to the dislocation slip plane, or form junctions \cite{terentyev2013transfer,bonny2016assessment}. After the junction formation, interstitial loops might be absorbed by dislocations \cite{bacon2006computer},  and the Burgers vector might be changed \cite{liu2008molecular}, or a relative stable junction is left. Generally, the resistant stress field induced by  irradiation defects, as well as direct junction formation can effectively stop gliding dislocations.  Thus, irradiation defects are more likely to inhibit strain bursts and dislocation avalanches. On the other hand, irradiation defects can be damaged or destroyed after the interaction with dislocations, thus reducing the resistance stress. Therefore, and depending on the dose and sample size, irradiation may promote or inhibit strain bursts, as observed experimentally and summarized in Table 1 of our work \cite{Cui2016irradiation}. Recent experiments on self-ion irradiated Ni pillars demonstrate that the stress drop magnitudes follows similar power law distribution (exponent -1.5). However, the upper limit of stress drop magnitude appear to be non-monotonic with the irradiation dose \cite{zhao2015situ}.  

\begin{figure}[htbp]
\centering
\includegraphics[width=.8\textwidth]{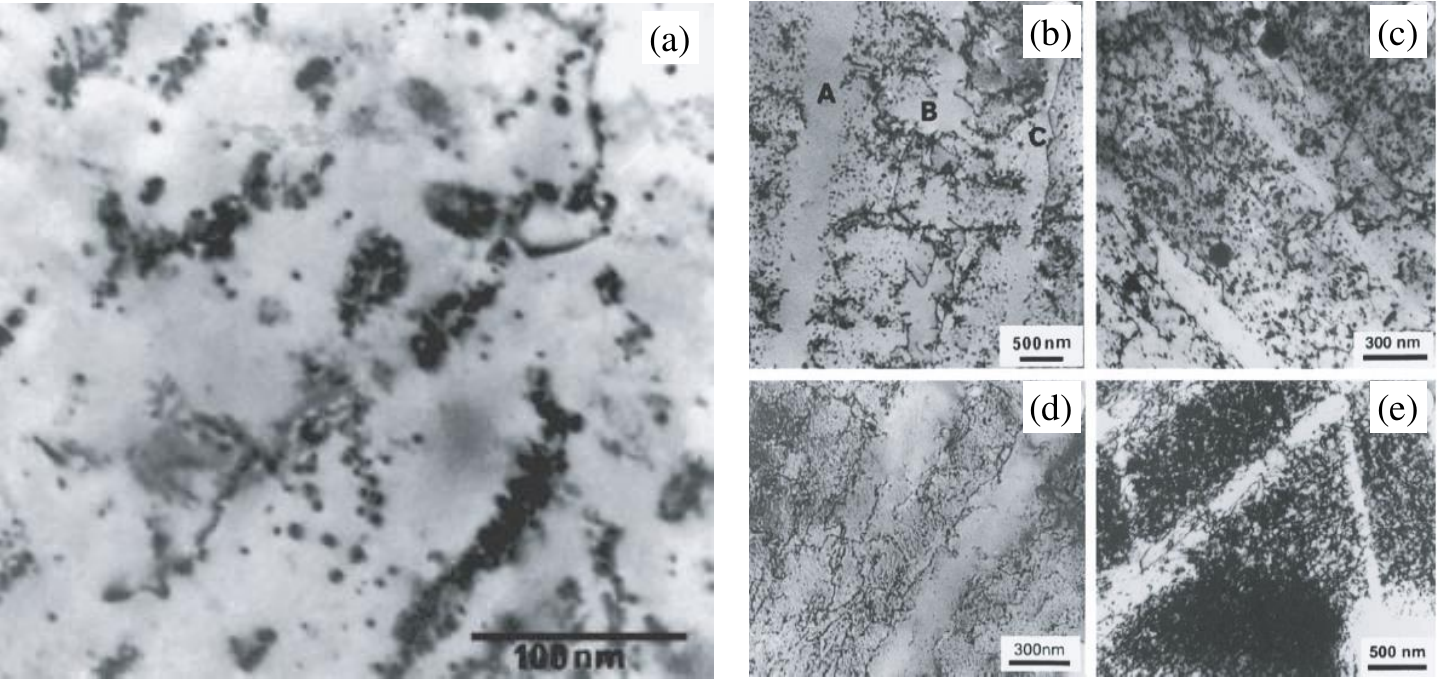}
\caption{(a) TEM micrographs showing rafts of loops formed in TZM (a molybdenum alloy containing 0.5\% Ti and 0.1\% Zr) during neutron irradiation at 350 $^\circ$C to a dose level of 0.16 dpa. (b-e) Some examples of cleared channel formation in neutron
irradiated metals: (b) single crystal Mo (50 $^\circ$C, $5.4 \times 10^{-4}$ dpa), (c) copper (100 $^\circ$C, $10^{-2}$ dpa),(d) copper (100 $^\circ$C, 0.3 dpa), (e) CuCrZr alloy (50$^\circ$C, 0.3 dpa). (Reprinted from \cite{Singh:2002JNM} with permission from Elsevier)} \label{fig:Singh-TZM-rafts-channels}
\end{figure}

The interaction of energetic neutrons with lattice atoms produces Self Interstitial Atom (SIA) clusters that
migrate rapidly in response to the internal stress field of dislocations.  They form what is known as \lq\lq rafts\rq\rq, as shown in Figure \ref{fig:Singh-TZM-rafts-channels}(a).  Some of these SIA clusters cling close to dislocations in what is known as \lq\lq decorations\rq\rq,  with the effect that dislocations become locked-in with reduced mobility or they become totally immobile. Once  external stress is applied, areas of stress concentrations or of mobile dislocations, start to respond. Strain is thus very localized and confined only to what is known as dislocation channels, as can be seen in Figures \ref{fig:Singh-TZM-rafts-channels}(b)--(e).

\subsection{Precipitation-Hardened Materials}
Experimental evidence demonstrated that serrated flow (abrupt stress drops) decreases or even disappears with aging in precipitation-hardened alloys, suggesting that precipitate coarsening inhibits or prevents the PLC effect \cite{chaturvedi1972serrated,pink2000serrated}. The explanation is that the solute content decreases with aging and that the stress field of precipitates inhibits the diffusion of solute atoms. On the other hand, and similar to the case of irradiation defects discussed above, small coherently ordered precipitates in precipitation-hardened alloys can be sheared off by gliding dislocations into fragments, become thermodynamically unstable and dissolve \cite{martin1997stability}).  This sequence of events leads to a decrease in the local flow stress and the occurrence of slip localization and softening \cite{Luft91}. Precipitate-free slip bands were observed in Al alloys and steels after monotonic and cyclic deformation. To avoid this inhomogeneous distribution of plastic slip, several possible techniques have been proposed.  These include changing from dislocation cutting of precipitates to a by-passing mechanism through particle coarsening, generating a microstructure with both coherent and incoherent precipitates, and adding homogeneously distributed hard inclusions \cite{Luft91}.  

In Ni$_3$Al alloys and superalloys, which contain coherent precipitates, intermittent plastic flow is observed when the diameter is below 40 $\mu$m \cite{dimiduk2007overview}. Below approximately 2 $\mu$m, similar jerky flow is also observed in Ni-based oxide-dispersion strengthened alloys (incoherent precipitate-strengthened alloy) \cite{girault2010strength}, and duralumin (aluminum 2025 alloy) micropillars \cite{gu2013size}. Note that the slip steps on these 1 $\mu$m duralumin is smaller and more homogeneously distributed, compared with the Al counterpart \cite{gu2013size}, which suggests smaller strain burst behavior with precipitation hardening in micropillars. Direct experimental evidence for suppressed strain bursts induced by precipitation are given in Al alloys with diameter ranging from 1 $\mu$m to 3.5 $\mu$m and Sc solute clusters or 3$\sim$8 nm Al$_3$Sc precipitates \cite{zhanga2016taming}. In addition, a transition from power-law scaling to a Gaussian distribution is observed for the detected strain bursts, by introducing such high pinning-strength disorder \cite{zhanga2016taming}. Gaussian burst distribution implies uncorrelated dislocation motion \cite{weiss2015mild}. However, there are also experimental observations indicating that strain bursts are promoted by introducing nanoscale $\eta '$ phase to Al micropillars \cite{hu2015stabilized}, because a number of dislocations are first trapped by precipitates, and then suddenly move in a correlated fashion, leading to large bursts. It was also observed that by introducing both second phase particles (nanoscale $\eta '$ phase) and grain boundaries (or coating interface) to ~200 nm Al pillars, strain bursts were effectively inhibited due to substantial dislocation storage and subsequent grain boundary mediated plasticity \cite{hu2015stabilized,li2017small}.

\subsection{Strain Bursts During Cyclic Loading}

 Discontinuous deformation during cyclic creep was reported in lead, Al alloys, and  steels, since the 1950s \cite{kennedy1956effect,kirk1977unusual}. Taking Cu-Al alloys as an example, the peak stress fluctuates by about multiple percent during cyclic straining. The occurrence of strain bursts is found not to be sensitive to the crystallographic orientation, but depends on the applied plastic strain amplitude \cite{wu2001cyclic,kaneko1997instability,kaneko1998cyclic}. No burst was detected at very large plastic strain amplitudes. Abel proposed that this is because the high stress at large strain amplitude makes it easy for dislocations to overcome obstacles \cite{abel1979low}. On the other hand, Kaneko stated that this is more possible due to the different dislocation structures at different strain amplitudes  \cite{kaneko1998cyclic,mori1979cyclic}. Moreover, each burst was found to be accompanied with the formation of significant slip bands in planar slip alloys (Cu-16at.\%Al) \cite{hong1990cyclic}. Cross slip was thought to play an important role in triggering sudden dislocation multiplication and strain bursts \cite{hong1990cyclic}. In addition, the interaction between alloying elements and dislocations, such as in dynamical strain aging, is also proposed to explain this burst behavior \cite{hu1992burst}.

In pure metals without alloying effects, strain bursts are also observed \cite{shin1988strain}. For example, strain bursts in bulk pure copper are reported during cyclic creep and cyclic tension/compression tests, which are explained through dislocation locking by the dislocation network and the subsequent sudden collapse of the locking mechanism. More specifically, it is proposed that loose dislocation walls may rearrange during cyclic deformation and suddenly collapse locally, or that the gliding dislocations within the cells suddenly penetrate towards dipolar walls \cite{lorenzo1982strain,lorenzo1984strain}. Afterwards, strain bursts are reported in Cu single crystals during the early stages of cyclic hardening. Since a stable microstructure is not expected during this stage, it is proposed that strain bursts originate from the formation and break-up of dislocation locks due to mutual interactions \cite{gong1996cyclic}.

\subsection{Strain Bursts and Surface Steps}

Intermittent strain bursts have been found to be associated with distributed surface steps. It was found that the slip patterns exhibit long-range spatial correlations, which is manifest in the appearance of fractal surface step features \cite{zaiser2001statistical}. In the 1980s, slip lines were observed to develop with a fractal structure in the uniaxial tensioned Cd single crystals \cite{spruvsil1985fractal}. Moreover, the fractal dimension of slip lines projected over an orthogonal segment was found to be 0.52 for microscale Cu single crystals \cite{kleiser1986fractal}. Correlation analysis carried out by Weiss and Marsen illustrates the fractal slip bursts pattern on macroscopic scales \cite{weiss2003three}. In addition, dislocation avalanche dynamics is spatio-temporally coupled: Namely, if two avalanches are close in time, they are also close in space. By monitoring the surface morphology evolution of plastically deformed Cu using atomic force microscopy and scanning white-light interferometry, Zaiser {\it et al.} revealed that the power-law correlations in the spatial distribution of plastic strain lead to self-affine surface profiles for scales ranging from 10 nm to 2 mm \cite{zaiser2004self}. The self-affine profile is quantitatively characterized by the Hurst exponent $H$. Taking the line profile $y(x)$ as an example, the average height difference $<|y(x)-y(x+L)|>$ changes like $L^H$. The fractal dimension of this line profile is calculated as $(2-H)$ \cite{zaiser2006randomness,kuznetsov2001fractal}. The fractal dimension of the corresponding surface is expected to be $(3-H)$. In deformed Cu polycrystals, the fractal dimension of the surface profile is found to increase from 2.0 to 2.3 with the first few percent of strain and then saturates \cite{zaiser2006randomness}. A similar trend is also observed in deformed Ni single and polycrystals \cite{meissner1998formation}. The surface profile analysis by Kuznetsov {\it et al.} suggested that the fractal dimension peaks before fracture \cite{kuznetsov2001fractal}. This was later confirmed by analytical considerations, which demonstrated that the peak fractal dimension of the bulk dislocation structure in deforming metals peaks at a certain strain close to the onset of necking\cite{vinogradov2012evolution,zaiser2006scale}.

\section{Models of Dislocation Avalanches and Strain Bursts}
\label{sec:model}
The wide-ranging experimental evidence of strain bursts, dislocation avalanches, and associated localized plastic flow, surface steps and cracks, prompted researchers to develop a variety of models. Such studies have been driven by several possibilities. First, it is important to quantify the noise in the plastic behavior for material design purposes, and consequently develop statistical approaches for dynamical investigations of microstructure evolution. Second, there is some evidence that abrupt plastic events are precursors to crack initiation, thus providing novel pathways towards building prognosis methods. Investigation of these events is the only pathway to developing self-consistent multiscale and continuum modeling approaches. Finally, there is a possibility that these abrupt events display a signature of SOC, and in this case, the origin is a combination of the extended character of dislocation defects and formation of metastable microstructures.

\subsection{Mutually-Independent Dislocations in Strong Disorder}
The primary cause for abrupt events in crystal plasticity originates in the very nature of dislocation dynamics; that is the jerky motion of dislocations. In small volumes, each dislocation movement is abrupt. The micromechanical character of such events has been characterized by various mechanisms, such as dislocation starvation, source exhaustion and junction saturation. Such events are dominated by extreme value statistics of the original material inhomogeneity and mirror the distribution of the underlying disorder. The simplest possibility for generalizing Orowan's law in order to include ``abrupt" features, is by considering the onset of plastic activity in crystals through the activation of mutually independent dislocation sources. This behavior is reminiscent of analogous behavior in fracture of beams: Consider a chain of $N$ beams with random and independent, identically distributed, failure strengths. Then, if the mathematical conditions of $\Phi(\sigma)=0$ holds for all $\sigma<\sigma_0$, and $\Phi(\sigma)\sim(\sigma-\sigma_0)^k$ for $\sigma\rw\sigma_0^+$ (equivalently, $\Phi(\sigma)$ is  the cumulative failure probability function, and is continuous with continuous derivatives across $\sigma_0$, where $\sigma_0$ is the minimum failure strength), then extreme value statistics  suggests that the cumulative extreme value distribution of failure probability as function of the applied stress $\sigma$ is the  Weibull distribution:
	\bea
	\Phi(\sigma)=\left\{ \begin{array}{rcl} 1 - e^{-(\frac{\sigma - \sigma_0}{\delta\sigma})^k}, \quad \textrm{for}  \quad  \sigma > \sigma_0 \\ 0, \quad \textrm{for}  \quad  \sigma \leq \sigma_0\end{array}\right.
	\eea
\noindent where $\delta \sigma \sim N^{-1/k}$ defines the range of the distribution. Here, $k$ is the Weibull exponent \cite{Weibull:1951bh}. 
	
Due to the primary configurational disorder distribution (defined by the initial dislocation microstructure or otherwise) there could be significant size effects in material parameters such as sample yield strength and hardening coefficient. If, for example, bulk pinned segments do not statistically form in small nanopillars, it is clearly expected that unconventional dislocation sources should prevail, such as single-arm dislocation sources and surface dislocation sources (where no pinned segments are required).

\subsection{The Pinning-Depinning Rubber Band Conceptual Model}

Another approach towards including abrupt crystal plastic events is through considering mutually independent dislocations that traverse a static forest of obstacles or/and junctions. Indeed, crystal plasticity in many cases, can be thought  of as the transport of dislocations in a medium that has static random heterogeneities (or, more commonly, ``quenched randomness") that apply forces on the dislocations and depend on their location. This phenomenon has been labeled as ``elastic interface depinning" \cite{Fisher:1998fk} and represents a general class of phenomena that may include interfaces between two fluids in a porous medium, domain walls in a random ferromagnetic alloy, vortex lattices in dirty type-II superconductors contact friction, and the motion of geological faults.

	The rubber band model is based on the concept of an interface between two phases that is driven by an applied force through an inhomogeneous medium, with the most crucial ingredients being: a) the elasticity of the interface caused by its interfacial tension, b) the interface is metastable and controlled by random heterogeneities, and c) the dynamical law of motion for the local interface position. If we assume over-damped interface dynamics for the interface displacement $u(\bfr)$, and typical elastic interactions, then there are some well-known basic results. First, if the driving force is small enough, the interface is immobile/pinned. Second, if the driving force increases slowly, it may overcome the pinning of a finite part of the interface, leading to an avalanche, stopped by the spatial heterogeneity. Third, if the force is large enough, the interface moves with some average velocity $\bar{v}$ via jerky motion in space and time. Fourth, for non-negative elastic interactions (such as simple springs), there is a unique critical force $F_c$ above which the mean interface velocity $\bar{v}(F)$ is non-zero. \cite{Narayan:1993ys}. For $F<F_c$, only avalanches may take place, and the correlation length $\xi$ can be defined as the distance beyond which avalanches are unlikely and if the maximum avalanche size is $S_0$, then $\xi\sim S_{0}^{1/2\alpha}$, where $S_0$ is the avalanche distribution cutoff
scale, $\alpha>0$ is another exponent. Furthermore, $\xi\sim 1/(F_c-F)^\nu$. In the mean-field approximation, it happens that $\nu=1/\alpha$. 

	Such exponent relations extend to the characteristics of slip events that take place for $F<F_c$. Namely, the event size distribution takes the form $P(S)=S^{-\tau}{\cal F}(S/S_0)$. (When $\tau$ is used as an exponent, it represents the typical critical exponent, instead of stress), where ${\cal F}(x)$ is typically modeled as an exponential function ${\cal F}(x)=\exp(-x)$,  $S_0=\xi^{d+\zeta}$, and $\zeta$ is the effective fractal exponent of the elastic interface. In other words, $\langle(u(\bfr) - u(\bfr'))^2\rangle\sim \xi^{2\zeta}$. This exponent of ``jerkiness" $\zeta$ is critical for the abrupt character of the elastic-interface motion, since it may be thought that the average linear length scale at which such abrupt events occur extend up to $\xi$. The relevant timescale would then be $\tau_\xi \sim \xi^z$, and the motion becomes smooth at larger space and longer time scales. The velocity of the elastic interface in this definition would be,
	\bea \label{interface_Velocity}
	\bar{v}\sim \frac{\xi^\zeta}{\tau_\xi} \sim (F-F_c)^\beta 
	\eea

	\begin{figure}[h!]
		\centering
	\includegraphics[width=0.45\textwidth]{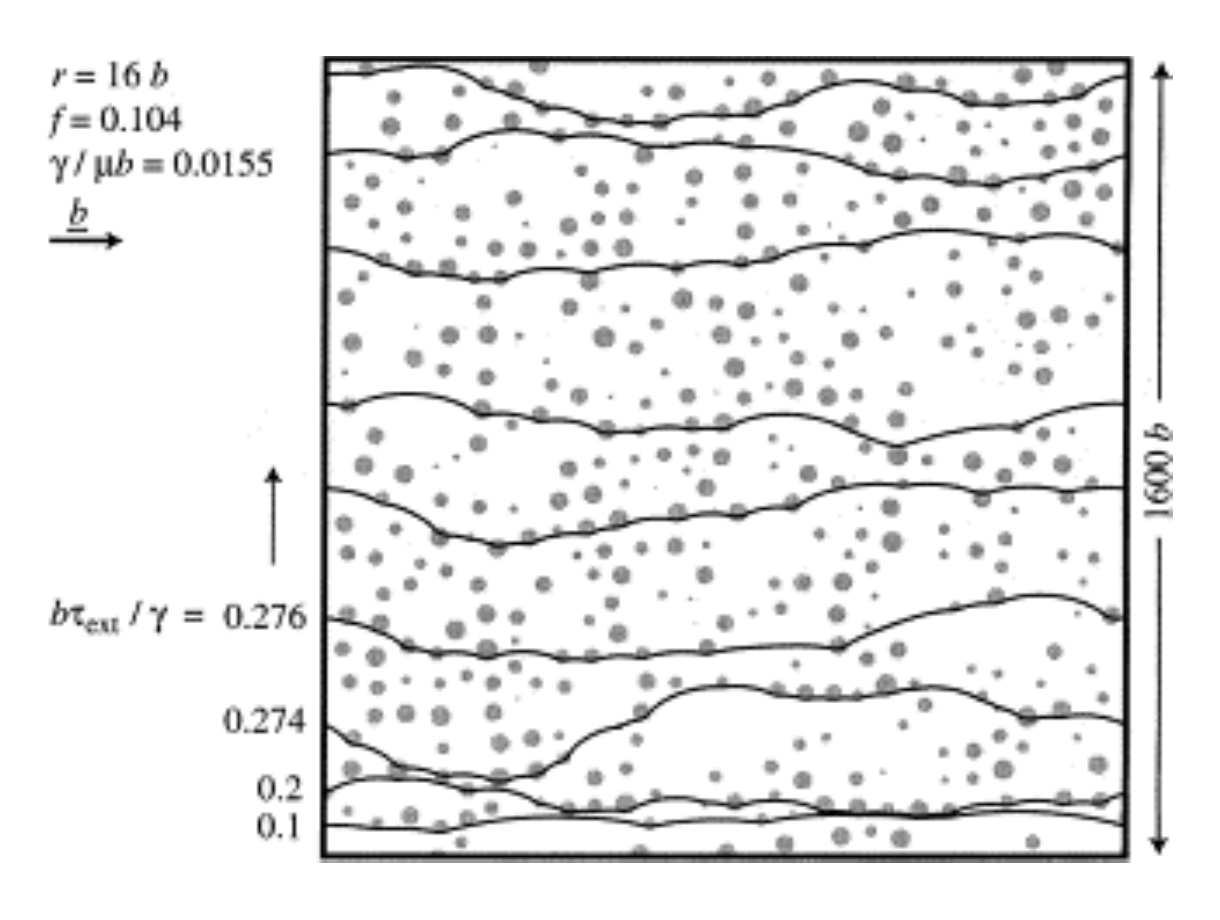}
	\includegraphics[width=0.45\textwidth]{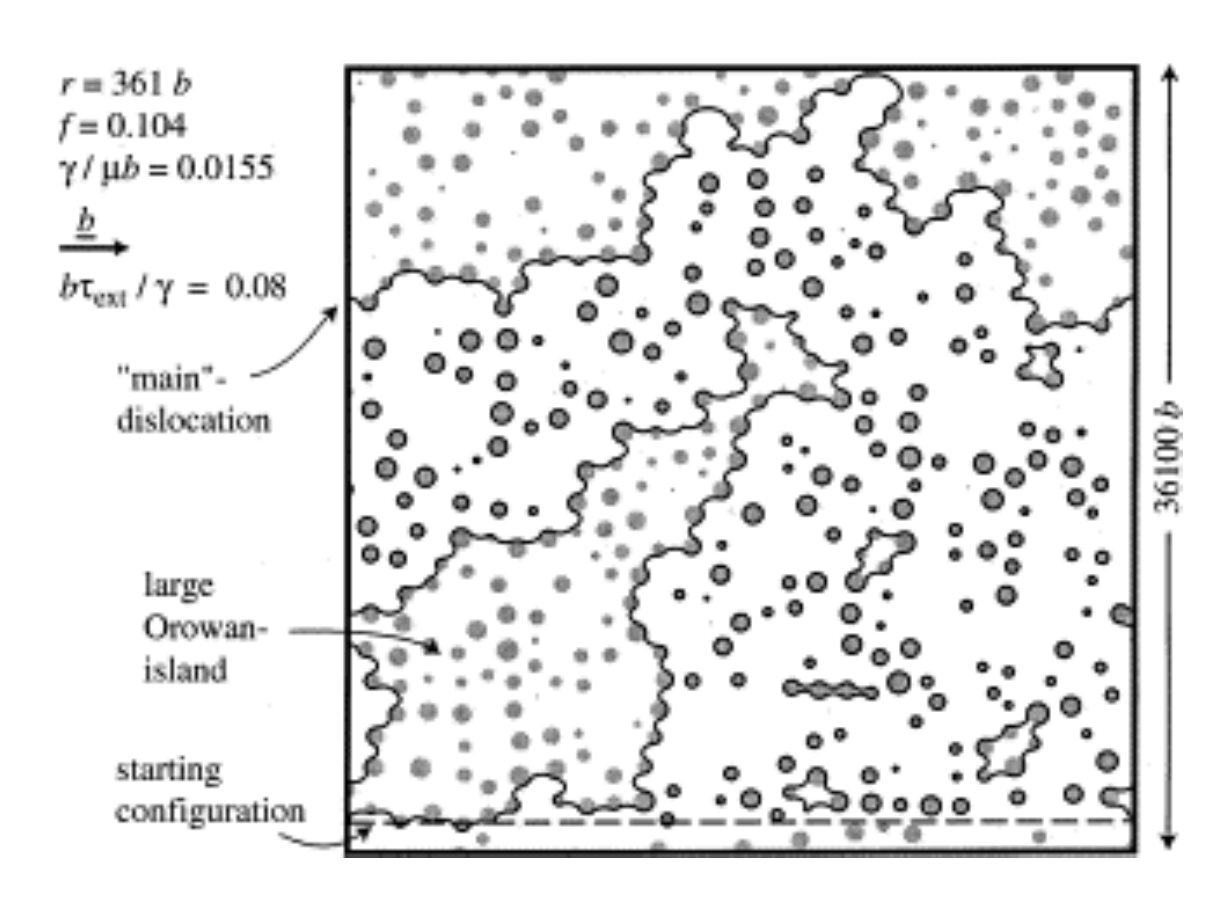}
	\caption{Simulated single dislocation in a field of small obstacles. The Burgers vector is $b$, the external shear stress $\tau_{\rm ext}$, the radius of the obstacles is $r$, at a fixed strain $\gamma$. Different lines correspond to different external shear stress. On the left, the obstacle radius is just $16b$ while on the right, it is very large at $361b$. (Reprinted from \cite{Mohles:1999zr} with permission from Elsevier)}
	\label{fig:disl-precip}
	\end{figure}
	
	The most natural example of elastic depinning theories on crystal plasticity is through the example of dislocation glide in precipitation hardened materials \cite{Mohles:1999zr}. As shown in Fig. \ref{fig:disl-precip}, and for small precipitate obstacles, a dislocation may glide in a similar fashion as in a system that undergoes elastic depinning. The dislocation is immobile or pinned for small applied stresses, while it acquires a slip velocity beyond $\tau_c$. The lowest three configurations are stable or pinned, while the other lines are in the ``moving" depinning phase.  The figure (right) shows a single dislocation in a field of large obstacles. Dislocations lose their monotonic character and the elastic interface ``breaks" into multiple lines, some of which might be ``critical", or in other words very close to a depinning stress: This interface multiplicity leads to critical stress ambiguity. such behavior may lead to ``self-organized" critical behavior, a behavior where there is always a set of elastic interfaces close to the critical depinning stress. While the analogy to elastic depinning is clear in this case, the situation drastically alters when obstacles become significantly larger, and the moving elastic interface develops large overhangs, a poorly studied regime in theories of depinning.

A description where such relations become exact is the so-called ``mean-field limit". In this case, one can show readily that  $\beta=\beta_{\rm MF}=1$ and $\nu=1/\alpha$, while $\zeta=0$\cite{Fisher:1998fk}. Such a limit can be realized if the {\it interaction} between two elements of the elastic interface is homogeneous, independent of their distance. While not physically realizable, this limit can provide a complete picture for the collective dynamics of the interface. Namely, if one assumes that the strain at the location of the dislocation on its glide plane is $\gamma(\bfr)$, then dislocation dynamics in the mean field limit (where the elastic, long-range interaction between $\bfr$ and $\bfr'$ is assumed to be just $\tau_{\rm int}(\bfr - \bfr')=A G (\gamma(\bfr) - \gamma(\bfr') )$, where $A$ is a constant and G is the shear modulus of the material) would take the form,
	\bea
	\frac{d \gamma(\bfr)}{dt} =M(A Gb(\gamma(\bfr) - \langle\gamma\rangle ) + \sigma_{\rm ext}b  - f_p(\bfr,\gamma(\bfr)))
	\label{eq:mf}
	\eea

\noindent where $M$ is the dislocation mobility, $f_p(\bfr,\gamma(\bfr))$ denotes the pinning/resistive forces originating in the precipitates, and $\langle\gamma\rangle = 1/L^2\int d^2 r \gamma(\bfr, t)$ denotes the spatial average of the dislocation location $\gamma(\bfr)$. $L$ is the characteristic length of the dislocation structure. $f_p(\bfr,\gamma(\bfr))$ denotes quenched noise in the sense that it is independent of time but renews whenever $\gamma(\bfr)$ changes appreciably. In this sense, the distribution of the variable $f_p$ is considered uncorrelated in the $\bfr$-space and $\gamma(\bfr)$-space. By taking the average of Eq.~\ref{eq:mf}, it is possible to show that the average position follows the stochastic differential equation \cite{dobrinevski2012nonstationary},
	\bea
	\partial_t \langle\gamma(t)\rangle = MF(\gav) - Mm^2 ( \gav - \gamma_{\rm ext}(t))
	\label{eq:abbm}
	\eea
 
\noindent where $m^2$ denotes a machine constant and $\gamma_{\rm ext}$ is the external driving strain rate in a presumed displacement-control mechanical test. The effective random force $F(\gav)$, which denotes the average random resistance force,  has been shown to be Gaussian with the correlations of Brownian motion,
	\bea
	\langle [F(x_1) - F(x_2)]^2\rangle = 2\sigma |x_1 - x_2|
	\eea
	where $\sigma>0$ characterizes the disorder strength. The mean-field model has been analyzed in great detail, due to analogies with other disordered systems, such as disordered magnets and earthquake faults, especially for the case of a constant driving external rate $\gamma_{\rm ext}(t)=\dot\gamma_{\rm ext} t$~ \cite{papanikolaou2011universality,alessandro1990domain}. The distributions of avalanche sizes and durations, as well as the mean shape of an avalanche pulse were obtained by mapping Eq.~\ref{eq:abbm} to a Fokker-Planck equation. It has also been shown that these results agree well with experimental systems described by long-range elastic interactions, such as thin-film permalloys~ \cite{papanikolaou2011universality} or in geological faults~\cite{Fisher:1998fk}. In particular, it may be shown that the average avalanche shape at binned duration $T$, defined as the probability of first return to the origin boundary condition, is~\cite{papanikolaou2011universality}:
	\bea
	\langle\dot{\gamma}(t)\rangle= \frac{4\sinh(t/2) \sinh(T/2 - t/2)}{\sinh(T/2)}
	%\dot\gav = \frac{4\sinh(t/2) \sinh(T/2 - t/2)}{\sinh(T/2)} + w[\sinh(T/2-t/2)/\sinh(T/2)]^2
	\eea
	The power spectral density $P(\omega)$ of the noisy signal can be defined as,
	\bea
	P(\omega) = \lim_{T\rw\infty} \frac{1}{T} \left | \int_{-T/2}^{T/2} e^{i\omega t} [ \langle\dot{\gamma}(t)\rangle - \langle\dot{\gamma}(t)\rangle_{time-aver}] dt\right|^{2}_{time-aver} 
	\eea
	where the ``time-aver" denotes temporal average quantities when it is mentioned. In particular, for a {\it stationary} signal where the only relevant variable is the time-difference of the signal, then (if we label $\gav$ as $\gamma_{av}$) then,
	\bea
	P(\omega)=\int_{-\infty}^{\infty} e^{i\omega t} \langle \dot\gamma_{av}(0)\dot\gamma_{av}(t)\rangle_{time-aver} dt
	\eea
	For the mean-field case and constant external driving rate, $\langle\dg(0)\dg(t)\rangle_{time-aver} = v e^{-|t|}$, here $v$ is $\dot\gamma_{\rm ext}$, and the power spectrum for the mean-field case is,
	$
	P(\omega) = \frac{2 v}{1+\omega^2}
	$, 
	namely, a Lorentzian. Moreover, the noise intensity distribution in the stationary regime is,
	$
	P(\dg) =  \frac{1}{\Gamma(v)} \dg^{v - 1} e^{-\dg}
	$. 
	The probability of an ``event" of  duration $T$ and size $S$, where the size of an avalanche is defined as $S=\int_{0}^\infty\langle\dot{\gamma}(t)\rangle dt$  and the duration as the first point when $\dg(T)=0$. $\Gamma(v)$ is the Gamma function at location $v$. Assuming a mock experiment where a finite driving strain is imposed at $t=0$, $W(t)=w\theta(t)$. Then, it may be shown that  the probability density for the avalanche duration is,
	\bea
	P(T) = \frac{\partial}{\partial t_0}\Big|_{t_0=T} P(\dg(t_0) = 0) = w exp[{- w/(e^T-1)}] / (2\sinh(T/2))^2
	\eea
	Giving $P(T)\sim T^{-2}$ for $T\ll1$, 
	\bea
	P(S) = \frac{w}{2\sqrt{\pi} S^{3/2}} \exp(-w^2/(4S) - S/4 + w/2)
	\eea 
It is remarkable in this model, that these finite-stress at $t=0$ event size distribution can be also calculated in the stationary regime with $W(t)=v t$, and then
\bea
P(S) = \frac{1}{\Gamma(\tau - 1) } \frac{1}{S} \left(\frac{S_0}{S}\right)^{\tau-1} e^{-S_0/S}
\eea
with $S_0=v_0^2/(4\sigma)$, and $v_0$ a small rate cutoff, necessary for the calculation to become possible.
Also,
$
\tau = \frac{3}{2} - \frac{m^2 v}{2\sigma},~ \mbox{and}~
S_0 = \frac{v_{0}^2}{4\sigma}
$
In this limit, by using the requirement that $P(S)dS = P(T)dT$, the duration distribution can be also calculated, giving $P(T) \sim T^{-(2-m^2v/(2\sigma))} $.

We should note that while criticality in equilibrium systems takes place in special points in parameter space, such as the liquid-gas critical point, there are more possibilities when out of equilibrium, such as SOC. Investigating {\it only} pristine crystals with very low initial dislocation densities, it is highly plausible that as additional plastic distortion is achieved, the system undergoes SOC, since only {\it some} of these dislocation segments might propagate according to their proximity to critical depinning.  Such a situation might emerge in the plastic response of a highly evolving, and possibly dense, dislocation forests.

In general elastic interface depinning considerations, the definitions for the exponents $\tau$, $\alpha$, $\nu$ and $\zeta$ are general  and can be defining possible ``universality classes", in analogy to the phase transition types (first/second order) for systems that display thermodynamic equilibrium~\cite{goldenfeld1992lectures}. For the case of crystal plasticity, there have been significant efforts to generalize the aforementioned mean-field results in various ways, with combinations of experimental data and coarse-grained models~\cite{Tsekenis:2013rr,Tsekenis:2011nx,Friedman:2012oq,Dahmen:2009kl,ispanovity2014avalanches,lehtinen2016glassy,papanikolaou2012quasi,Laurson:2013ff}.

\subsection{Two-Dimensional  (2-D) Dislocation Dynamics Models}

Dislocation dynamics simulations in 2-and 3-dimensions have provided a wealth of evidence that depinning theories may apply to the description of some aspects of crystal plasticity. Initial efforts appeared in simulations of gliding edge dislocations that were initiated in a random ensemble, in the absence of obstacles/precipitates and Frank-Read sources. Such an ensemble may be applicable in highly prestrained thin films, but it should be mainly viewed as a toy model where explicit dislocation mechanisms may be tested. After a decade of research, the identification of this 2-D, plastic behavior as elastic interface depinning has been concluded to be insufficient \cite{ispanovity2014avalanches}. The primary reason for this failure is clearly the absence of explicit obstacles, thus the behavior resembles more that of amorphous solids
\cite{talamali2011avalanches,Eshelby:57,baret2002extremal,vandembroucq2004universal}.  An interesting dislocation mechanics problem that may describe collective dislocation phenomena, and that can be addressed both in theory and simulations, is the loading of samples that contain only edge dislocations   in a periodic potential \cite{derlet2013micro} or in an initial random configuration \cite{groma1997link,zaiser2001statistical}. If only edge dislocations exist in the system, a 2-D representation suffices. While isolated edge dislocations and their theoretical description may find application only in the context of thin films \cite{xiang2005plane,xiang2006bauschinger,davoudi2014bauschinger,shishvan2010bauschinger}, it is a simple enough problem that can provide insights into complex 3-D dislocation dynamics as well as continuum theories of dislocations \cite{yefimov2004comparison}. 

A basic version of 2-D dislocation dynamics that has been used for describing the non-smooth avalanche behavior at the microscale is based on a randomly ``mixed" ensemble of edge dislocations \cite{groma2003spatial}. The system consists of parallel edge dislocations that lie on parallel slip planes, in a single slip system. If one denotes the position of the $i$th dislocation as $\bfr_i=(x_i,y_i)$ and its Burgers vector is $\bfb_i=s_i(b,0)$ where $s_i=\pm 1$ is the ``charge" sign, the equation of motion for a single dislocation is,
\bea
\dot x_i = M s_i b \left[ \sum_{j=1,j\neq i}^N s_j \tau_{\rm ind} (\bfr_i - \bfr_j) + \tau_{\rm ext} (\bfr_i)\right],\quad
\dot y_i = 0
\eea
In the case of edge dislocations, the shear stress field generated by each dislocation is exactly known $\tau_{\rm ind}=\frac{\cos(\phi)\cos(2\phi)}{r}$ with $\phi=\tan^{-1}(y/x)$, improving remarkably the speed of simulations. Furthermore, $\tau_{\rm ext}$ denotes the external shear stress and $N$ is the total number of dislocations. Equal number ($N$/2) of positive and negative dislocations are initiated in the system and then, the initial position of each dislocation is chosen randomly with the configuration relaxed to a mechanically equilibrated configuration. 

Given the scale-free character of the random initial condition, it is natural to assume length, time and stress scales are mainly dependent on the dislocation density $\rho$: $l_0=1/\sqrt{\rho}$, $\tau_0=1/(\rho M G b^2)$ and $\sigma_0=G b \sqrt{\rho}$. It is important to also notice that the dynamics of the dislocations is over-damped and also highly constrained, since they may only move along the x-direction. Thus, if initialized randomly, the system remains in a state of quenched disorder with a high degree of metastability, naturally displaying frustrated glass-like features in its dynamics \cite{bako2007dislocation}.
 
It is natural to ask the question in the context of the the 2D random edge dislocation ensemble model: how close is the statistical event dynamical behavior to elastic interface depinning? The answer to this question can be based on investigating the crossover of the collective dynamics as one introduces precipitates explicitly. In the presence of precipitates, depinning should dominate the mechanical behavior, however in a pristine crystal, clearly the frustration of the dislocation network originates in ``intrinsic" obstacles that are generated through the complex temporal and spatial evolution of the dislocation ensemble, as more dislocations nucleate in the material. Recently, Ovaska {\it et al.} \cite{ovaska2015quenched} proposed a way to do so, through generalizing the standard 2D-DDD model to include a random configuration of $N_s$ quenched pinning centers, as it could be generated by low mobility solute atoms. These immobile solute atoms interact with dislocations via short-range elastic interactions derived from non-local elasticity considerations \cite{wang1990non}. However, there are ``intrinsic" effects that cannot be included in 2D-DDD for topological reasons, such as the inclusion of dislocation junctions (as observed in three-dimensional simulations). The role of the ``intrinsic" character of such junction-obstacles into crystal plasticity behavior is evident in the fundamental differences between the mechanical behavior in crystals that progressively harden in comparison to crystals that have increasing amounts of precipitates. However, in theory, the fundamental differences of such ``intrinsic" junction-obstacles and their role into the plastic flow have not been explored. In contrast, such a question has been recently explored in sheared yield stress fluids \cite{lerner2012unified,puosi2015probing}.
	
The principal difference between ``intrinsic" (junctions, dipolar bound states) and extrinsic (precipitates) obstacles in yield stress fluids and other amorphous materials is in the behavior of the local yield stress distribution in the avalanching steady state. While there is not yet clear conclusion, it appears that the local yield distribution has ``fat" tails, resembling a Weibull distribution. The $k$ exponent in the Weibull distribution  is known to influence the behavior of strain bursts, since it clearly describes the propensity to further continuation of an avalanche.

The benefit of the random 2D-DDD edge model is that it is possible to build a continuum description, using mean-field density distributions, similar to the equilibrium classical many-body theory. Basic statistical mechanics theory provides the logical connection between the spatial distribution functions of all orders and the basic interparticle interactions that operate in the material system of interest \cite{hill1956tl,hansen1976theory}. In order to reduce the formal theory to a tractable form that is capable of producing numerical results for model systems, one performs {\it ad hoc} closure approximations \cite{kirkwood1935statistical} that may be later justified by explicit microscopic simulations or experiments \cite{kirkwood1942radial,percus1958analysis,verlet1964theory}. It has become clear how to construct both local (such as back-stress) and non-local (long-range stress) continuum terms of the dislocation density evolution \cite{groma2016dislocation,groma1997link,zaiser2001statistical,groma2003spatial,groma2007dynamics}.  

Finally, we should note that the random-edge 2D-DDD model contains no dislocation sources and that it is typically investigated in periodic systems.  Thus, it is not possible to study the progression of avalanche behavior as the dislocation density increases in finite samples, where dislocation sources can be either bulk --typically of the Frank-Read type-- or unconventional, such as surface and single-arm sources \cite{greer2011plasticity}. Moreover, sources and obstacles develop a strong interplay that involves dislocation pile-up formation. Chakravarty {\it et al.} \cite{chakravarthy2010effect} have developed a direct connection between the yield strength of a system of edge dislocations in single slip and obstacle spacing, obstacle strength, source nucleation strength and average source spacing.  Avalanche behavior and its connection to strengthening in 2D-DDD systems with sources has been investigated in Ref. \cite{Papanikolaou:2015wt}, where a minimal 2D-DDD model with sources and obstacles, is constructed; This model captures the basic aspects of uniaxial compression/tension of nanopillars, including yield strength size effects as well as stochastic effects of post-yield plastic flow. 

\subsection{Continuum Dislocation Models and  the Complexity of Junction Formation and Multiplication}	

Generalizing idealized 2D models into 3D DDD and 3D continuum descriptions is a highly non-trivial task that is challenged by various modeling complexities: i) topological complexity of dislocation ensembles, ii) slow temporal effects such as aging, iii) multi-physics aspects, such as vacancy/solute diffusion, iv) slip system projection of dynamics. In order to include such complex effects in a minimal way, there have been various models and approaches. In connection with 2D models, there have been efforts by El Azab and collaborators \cite{el2000statistical,fivel1999linking,xia2015computational,el2006statistical} to generalize the 2D closure approximations and construct general dynamical evolution equations for all dislocation components. These models have been used to study spatial self-organization and provided many details of the collective interaction between dislocation ensembles. However, it is a difficult and daunting task, both numerically and experimentally, to analyze spatial patterns and identify the variety of effects related to avalanche behavior in crystal plasticity: The combination of abrupt events, stochasticity and pattern formation have been slightly pursued using a variety of continuum approaches that are not yet well grasped, using: i) phase-field atomic density variables~\cite{Chan:2010tg} with various artifacts due to extreme loading rates, ii) phase-field dislocation density variables, but using very simplified dynamics~\cite{Koslowski:2004fy,Koslowski:2007la}  iii) the Nye dislocation density tensor with various approximations to dynamics laws~\cite{Fressengeas:2009jt,chen2010bending,chen2013scaling}. Moreover, all these methods are strongly influenced by numerical solution issues, given the extreme timescale separation between the dynamics during the abrupt plastic event timescale and the one during overall loading.

Another framework that was inspired by the many successes of the field of non-linear dynamics in explaining spatio-temporal self-organization has been advocated as a convenient framework for description of the collective behavior of dislocations.  Instead of representing the equations of motion of each dislocation line and describe its interactions with other dislocations or with obstacles to its motion, dislocations are viewed as ``particles" that can interact and obey conservation laws. In other words, each dislocation is described as a chemical species in a medium, where it can collide, react, and produce by-products.
In a first attempt, Kubin and co-workers
 \cite{kubinestrin-90,kubinestrin-92} described jerky flow with two weakly nonlinear kinetic
equations describing the behavior of two dislocation populations,
the mobile one, of density $\rho_m$, and a forest population, of
density $\rho_f$:

\begin{eqnarray}
\partial_t \rho_m &=& \frac{C_1}{b^2} - C_2\rho_m -\frac{C_3}{b}\rho_f^{1/2}\nonumber\\
\partial_t \rho_f &=& C_2\rho_m +\frac{C_3}{b}\rho_f^{1/2} - C_4\rho_f
\end{eqnarray}
\noindent where $b$ is the modulus of the
Burgers vector, $C_1$ is the generation rate constant of mobile dislocations,
$C_2$ the immobilization rate constant,  $C_3$ the storage rate constant through interaction with forest dislocations, and $C_4$ is the recovery rate constant.

This model is able to predict the critical conditions for the
PLC effect and to determine the strain rate intervals in
which jerky flow can be observed. In experimental systems such as
Cu-Mn, Al-Mg, Cu-Zn, Au-Cu \cite{kubinestrin-92}, a good
agreement with experimental data has been obtained. Although the
model is based on essential ingredients such as annihilation,
trapping and interactions between dislocations, it is yet
oversimplified. A criticism of this simple model is that it is structurally unstable,
{\it i.e.} it does exhibit oscillations, but their periods and
amplitudes are determined by the initial conditions, and not by the
properties of the nonlinear system itself. Since jerky flow is a
manifestation of intrinsic material behavior, and corresponds
to self-sustained oscillations associated to a limit cycle, its
frequency and amplitude should be determined by the properties of
the dynamics and not by initial conditions.

According to the theory of nonlinear oscillations, it is
impossible to have a limit cycle surrounding an unstable node, or
focus, in a system of differential equations with only two
variables, if the nonlinearities are quadratic or lower. The model
has thus to incorporate other mechanisms. By considering more dynamical variables,
which increases the number of equations to three or more, would
allow us to retain only quadratic nonlinearities as
dominant ones. This approach has been proposed by Ananthakrishna
\textit{et al.} \cite{anantha-82,anantha-93}, which
satisfies all of the criteria discussed above, and satisfactorily
reproduces jerky flows and strain bursts.

In addition to mobile and forest dislocations, this model may also
consider dislocations surrounded by solute atoms, which are much
slower than the mobile ones, and mimic the phenomenology of dynamic strain
aging. Dynamic strain aging can be modeled in various ways~\cite{Rizzi:2004fu,Hahner:2003lh,anantha-82,anantha-88} and here, we display basic details in the context of the model in Refs.~\cite{anantha-82,anantha-88}: If  the density of solute atom - surrounded dislocations is represented by $\rho_s$, then the aforementioned rate equations for various dislocation densities may be generalized:
\begin{eqnarray}\label{ananthakrishna}
\partial_t \rho_m &=& \theta v_g(\sigma )\rho_m -\mu \rho_m^2 - \mu\rho_m\rho_f +\lambda\rho_f -\alpha\rho_m\nonumber\\
\partial_t \rho_f &=& k\mu \rho_m^2 - \mu\rho_m\rho_f -\lambda\rho_f +\beta\rho_s\nonumber\\
\partial_t \rho_s &=& \alpha\rho_m -\beta\rho_s
\end{eqnarray}
\noindent where the first term on the right-hand side of
the first equation describes dislocation generation by multiple
cross-glide or Frank--Read mechanisms.~$\theta$ is a
kinetic rate constant and $v_g$ is the average glide velocity of
mobile dislocations. The second term represents the pair
annihilation of mobile dislocations (with rate constant
$(1-k)\mu$) and their transformation into forest dislocations
(with rate constant $k\mu$). The third term corresponds to the
annihilation of a mobile dislocation with an immobile one, and the
last two terms represent the freeing of forest dislocations and
the transformation of mobile dislocations into solute surrounded
dislocations, respectively. These terms have their counterpart in
the other kinetic equations. Furthermore, the $\beta\rho_s$
contribution expresses the fact that, when the solute cloud
increases around a mobile dislocation, it eventually becomes
immobile, and contributes to the forest. The system
(\ref{ananthakrishna}) is formed by a set of coupled nonlinear
ordinary differential equations, and it has been shown that, in
some well-defined parameter range, it admits oscillatory solutions
of the limit cycle type, induced by Hopf bifurcation. The model
(\ref{ananthakrishna}) may finally be coupled to the machine
equation describing the load sensed by the load cell, namely:
\begin{equation}
 \dot \sigma_a = \kappa \lbrack \dot\epsilon - bv_g(\sigma_e)\rho_m\rbrack
\end{equation}
where $\dot\epsilon$ is the applied strain rate, $\kappa$ is the
effective compliance. The second term at the right-hand
side is the plastic strain rate, given by Orowan's law, where
$\sigma_e$ is the effective stress, with $\sigma_e = \sigma_a -
c\rho_f^{1/2}$. If one, furthermore, assume that the glide
velocity follows the phenomenological law $v_g =
v_0(\sigma_e/\sigma_0)^m$, where $\sigma_0$ is the yield stress, $m$ is constant, 
the outcome of the model may be related to experimental data.

This model has been shown to successfully reproduce the classical
PLC effect, and some features of temporal instabilities in fatigue and
strain bursts in ramp loading of metallic specimens
 \cite{Neumann,glazov-95}. It is interesting to note that, as many dynamical models of this
type, which may be encountered in nonlinear physics, this model
also exhibits chaotic solutions in
certain parameter range. In fact, it presents an infinite sequence
of period-doubling bifurcations ultimately leading to
chaos. Although the strain rate interval where
chaos is predicted is much smaller than the domain
where limit cycles are observed, chaotic plastic flow has nevertheless be observed experimentally
 \cite{Noronha-97,neuhauser-93}.

It thus appears that nonlinear dynamics is able to model plastic
instabilities and to describe various complex stress or strain
behavior which are experimentally observed. This modeling is
based on rate equations describing the evolution of dislocation
densities as the result of their nonlinear interactions. Up to
now, the spatial aspects of plastic instabilities have not been
considered, since the modeling described in this section is
supposed to be valid in a cross-section of the specimen, with no
coupling between different spatial locations. However, such couplings
are at the origin of spatio-temporal phenomena such as propagative
localization, L\"uders band propagation, persistent slip band and
dislocation microstructure formation. 

Spatial aspects have been considered by Walgraef and Aifantis \cite{walgraef1985dislocation}, where they studied the spatial self-organization of dislocations in various patterns (e.g. ladder structures, planar arrays, and dislocation cells).  Considering systems oriented for
single slip, the conservation equations for the forest ($\rho_s$) and mobile ($\rho_m$) densities read (with ${\bf v}_g = v_g {\bf 1}_x$ ):
\begin{eqnarray}\label{dislocdyn}\partial_t\rho_s &=& D_s\triangle \rho_s + c - v_sd_c\rho_s^2 - \beta\rho_s +v_g G(\rho_s)\rho_m \nonumber\\
\partial_t\rho_m^+ &=& - {\nabla}\cdot v_g\rho_m^+ +{\beta\over 2}\rho_s - v_g G(\rho_s)\rho_m^+
\nonumber\\
\partial_t\rho_m^- &=& {\nabla}\cdot v_g\rho_m^- +{\beta\over 2}\rho_s - v_g G(\rho_s)\rho_m^- \end{eqnarray}
The Walgraef-Aifantis model includes spatial gradients in the densities via a ``diffusion-like" term in the $\rho_s$ equation ($D_s\triangle \rho_s $), and drift-type terms in the two $\rho_m$ equations ($\nabla\cdot v_g\rho_m^{\pm}$).  The model has been successsful in predicting the Persistent Slip Band ladder structure under cyclic loading, and various aspects of spatial patterning under monotonic loading conditions.  Nevertheless, the model has not been utilized to study the temporal behavior of dislocation avalanches and the associated strain bursts.

\subsection{Direct Numerical Simulations with Discrete Dislocation Dynamics}
\begin{figure}
\centering
\includegraphics[width=.95\linewidth]{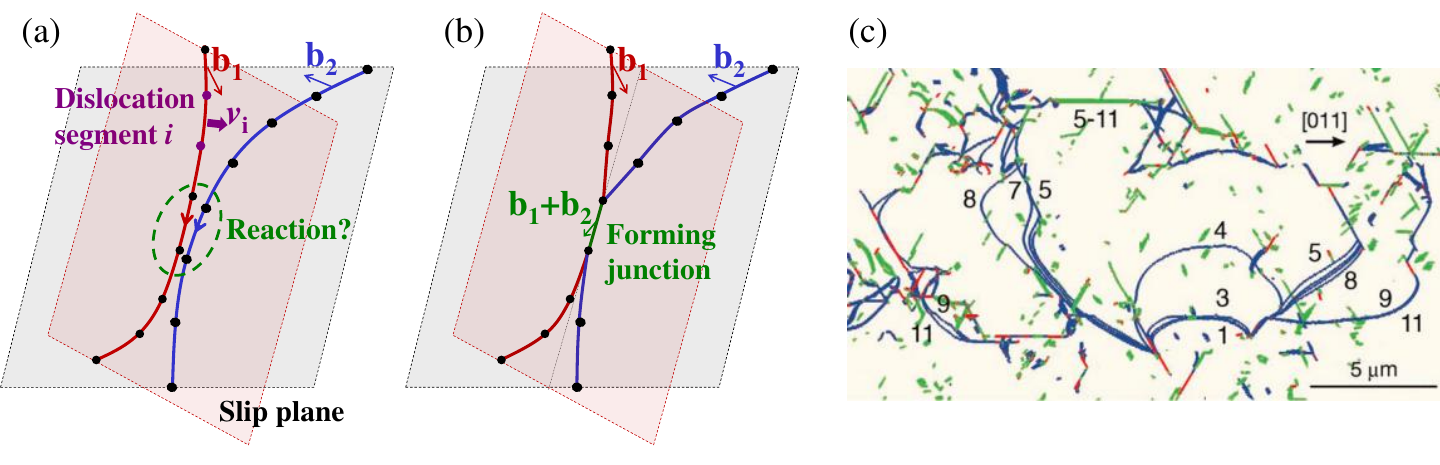}
\caption{(a)Schematic showing discrete three dimensional dislocation lines, their motion and their reactions; $b$ and $v$ represent the burgers vector and velocity; (b) Schematic showing two dislocation lines form one type of junction when $b_1^2+b_2^2>(b_1+b_2)^2$; (c) One example showing the interaction of gliding dislocations in a pre-designed dislocation forest (Reprinted from \cite{devincre2008dislocation} with permission from The American Association for the Advancement of Science) }
\label{fig:3ddd}
\end{figure}

Three dimensional discrete dislocation dynamics (3D-DDD) simulations, directly treating dislocations as the basic degrees of freedom, have been developed to quantitatively study the dynamic evolution of realistic dislocation structures \cite{bulatov2006computer,kubin2013dislocations,po2014recent}. An arbitrary dislocation is discretized into straight or contiguous curved spline  segments \cite{ghoniem2000parametric}, which contain the information of slip plane and Burgers vector, as shown in Fig. \ref{fig:3ddd} (a). Dislocation mobility laws for a specific crystal structure are determined based on the collective information from lower-scale molecular dynamics simulations, experimental results, as well as analytical theory \cite{kubin1998mesoscopic,cai2004mobility,po2016phenomenological}. The velocity of each dislocation segment (such as $v_i$ in Fig. \ref{fig:3ddd}(a)) is a function of the applied stress, the image force induced by specimen surfaces, the long-range interaction stress induced by other dislocations, temperature, and the nature of the Burgers vector orienation.

The topology is frequently updated to deal with surface annihilation and the short-range interactions between dislocations, such as collinear interactions, the formation and destruction of  junctions, etc. When two dislocations are close enough, a reaction may occur if it is energetically favored (see Fig. \ref{fig:3ddd}(a) and (b)). The reaction type is determined by the relationship of the Burgers vector and slip planes of interacting segments \cite{groh2009advances}. Several types of dislocation junctions have been found to have different strengths \cite{devincre2008dislocation}. Note that very strong junctions are sometimes artificially introduced by placing Frank-Read sources with indestructible pinning points. This will result in an  overestimate of the source strength, and an artificial increase in dislocation density \cite{lee2013emergence}. Therefore, the natural consideration of the formation and destruction process of junctions requires careful treatment of the initial dislocation configurations \cite{cui2014theoretical}. Stress-free relaxation of randomly distributed dislocations is required to obtain a stable initial dislocation configuration. Through coupling with the finite element method, based on the superposition method \cite{van1995discrete,po2014recent} or the eigen strain method \cite{lemarchand2001homogenization,cui2015quantitative}, 3D-DDD is now capable of tackling problems with complex boundary conditions, free surface effects, as well the influence of finite deformation.
 
3D-DDD simulations have clear advantages over the computationally more efficient 2D-DDD method, although at the cost of computational complexity.  In 2D simulations, only edge dislocations are considered, which makes it difficult to model plastic deformation in BCC metals, where screw dislocations play a dominant role. Obstacles with specific strengths are introduced in 2D-DDD to approximate the barrier effect of forest dislocations. For a realistic dislocation system, however, forest dislocations generate long-range stress fields, which would be difficult to be captured as simple obstacles. Moreover, forest dislocations evolve dynamically through the formation of dipolar loops and jogs, as well as the formation and destruction of different kinds of junctions etc (see Fig. \ref{fig:3ddd}(c)).~Therefore, 3D-DDD can capture naturally the dynamical evolution of dislocation networks and the correlated dislocation motion with minimum ad hoc assumptions.  Additionally, the formation of dipolar-loops through several cross-slip mechanisms has been found to play an important role in intermittent plastic deformation \cite{crosby2015origin}. However, such thermally activated cross slip process of screw dislocations cannot be taken into account in 2D-DDD simulations.

\section{An Overview of Modeling Results}
\label{sec:results}

\subsection{Weak Link Statistics: The Usual Suspect}

Weibull statistics is a general behavior that typically emerges as the distribution of the critical stress when independently failing/yielding entities are sampled. It is for this reason that it provides a valid explanation for plastic yielding when the initial dislocation density is low. Weibull statistics may thus provide an explanation of the critical stress for nano-indentation. Support for Weibull statistics in the stress levels required for strain bursts has been provided recently by Maass and Derlet \cite{derlet2015probabilistic}. Simulation evidence of Weibull statistics has been clearly observed in 2D- and 3D-DDD simulations~\cite{Ispanovity:2013dq}. Additionally, there has been experimental evidence in the context of micropillar deformation. Norfleet\etal~\cite{Norfleet:2008cr} suggested this mechanism through TEM investigations of deformed Ni pillars with diameters $1-20\mu m$, with further corroboration by 3D DDD simulations~\cite{El-Awady:2009nx,Senger:2011oq}. Further experimental evidence was provided in Ref.~\cite{Rinaldi:2008kl} on nickel nanocrystalline nanopillars with diameters $160\pm30$nm and Weibull exponent $k\in(3.2,5)$. In the study of Senger \etal \cite{Senger:2011oq}, a thorough analysis of the flow stress values was performed  for a variety of pillar sizes, crystalline orientations and aspect ratios, showing that they follow a Weibull distribution with $k\in(4,20)$. The ultimate reason for the effect in most cases is the existence, in small pillars, of few, almost independent, Frank-Read sources, that activate when the resolved shear stress surpasses the weakest source activation stress. This is exactly the condition for generating Weibull statistics. However, the observation of Weibull distributions in nanocrystalline samples suggests that there are also collective dynamical effects that are driven by {\it similar} distributions, while their fundamental origin is clearly non-trivial \cite{Ispanovity:2013dq}. 

\subsection{Results of Interface Depinning and Mean-Field Models}

Interface depinning models can be very useful towards providing microscopic insights and explanations for multiscale self-organization phenomena in crystal plasticity that involve the competition of short (abrupt avalanches) and long (aging, creep) timescales: Clearly, the oldest concept in such self-organization of systems under shear is stick-slip dynamics, where the sequence of stick and slip phases in a contact, determine the overall resistance in sliding friction. The mechanical energy dissipates in the sudden slip phase, while the stick phase is characterized by contact strengthening mechanisms, contact aging. Recently, it has become clear that stick-slip dynamics is the outcome of this competition between contact aging and energy dissipation during slip. The simplest way of demonstration is a variant of the Burridge-Knopoff (BK) spring-block model~\cite{Braun:2002sf} of earthquakes~\cite{Burridge:1967cs}, similar to the model that was studied by Olami, Feder and Christensen (OFC)~\cite{Olami:1992jt}, where contact aging is phenomenologically modeled~\cite{Braun:2002sf}. While not the only way to induce stick-slip transition in a model~\cite{Carlson:1994rw}, the role of contact aging becomes clear when it is considered that the model that includes it predicts realistic values for the stick-slip transition displacement-rate. A natural outcome of that model is the suggestion that the ``smoothness" of sliding at the macroscale gives its place to microscopic stick-slip behavior.

An analogue of the Braun and R\"{o}der model in crystal plasticity was only suggested recently~\cite{papanikolaou2012quasi}, in order to explain the onset of stick-slip behavior during the uniaxial compression of $20\mu$m single-crystal $Ni$ micropillars with a decreasing displacement rate. The analogue of the contact aging in frictional stick-slip modeling was identified in terms of aging of dislocation junctions, which could originate in cross-slip and double cross-slip dislocation mechanisms~\cite{Puschl:2002cl}. However, assuming the generality of the phenomenon in any OFC-like model where dislocation elasticity and aging are combined \cite{Jagla:2007vf,papanikolaou2012quasi,Braun:2013by,Jagla:2010fq} any kind of dislocation-junction or precipitate aging (such as dynamic strain aging \cite{cottrell1949dislocation} would cause an analogous stick-slip transition to emerge as the displacement rate is decreased in crystalline micropillars with significant aging rate. It is simple to build a discrete dislocation model that is a direct analogue of the OFC friction model construction.  Furthermore, the phenomenology of this model directly corresponds to the model of Ref.~\cite{Braun:2002sf} if it is assumed that each junction or obstacle stress $f^s_i(t)$ displays aging:
	\begin{equation}
	f^s_i(t)=f^s_{min} + (f^s_{max}-f^s_{min}) \left[ 1 - e^{-t / \tau} \right]
	\end{equation}
where $\tau$ is the characteristic junction/obstacle relaxation timescale, assumed to be homogeneous across all junctions, $f^s_{min}$ is the starting force for each junction and $f^s_{max}$ is the maximum force at which the junction stress saturates.

		\begin{figure}[h!]
		\centering
\includegraphics[width=.8\linewidth]{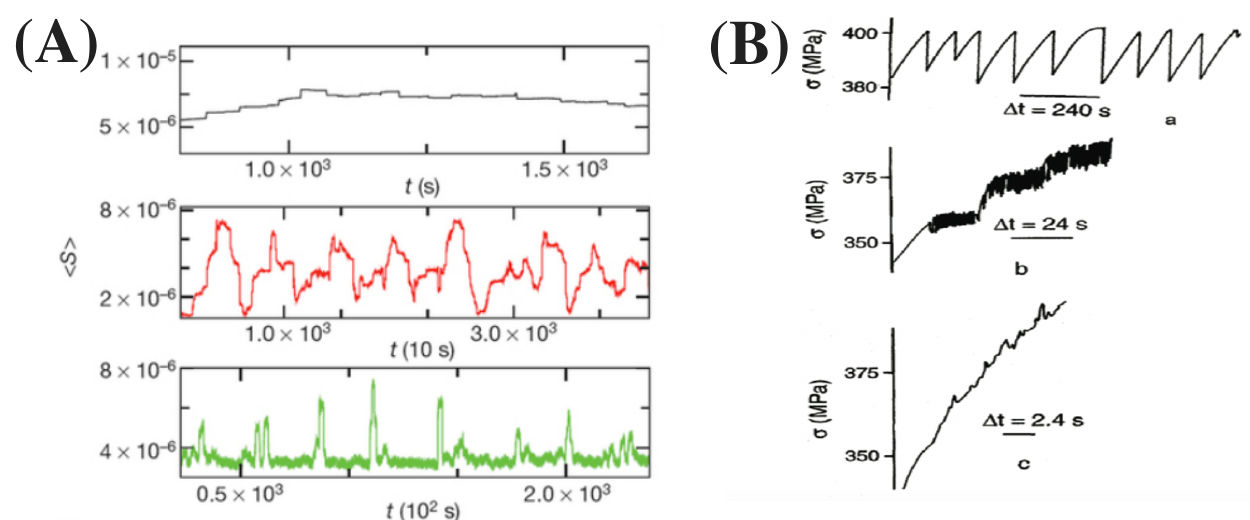}
\caption{(A) Stick-slip oscillatory phenomena in crystal plasticity of $Ni$ $20\mu$m micropillars at nominal strain rates $10^{-4}$s$^{-1}$(top), $10^{-5}$s$^{-1}$, $10^{-6}$s$^{-1}$(bottom). Locally averaged strain-rate vs. time is shown. (Reprinted from \cite{papanikolaou2012quasi} with permission from Nature Publishing Group) (B) Stress vs. time curves for Al$-$5at\%Mg alloy at T = 300K. From top to bottom, the strain rate is $5\times 10^{-6}$s$^{-1}$ (Type C), $5\times 10^{-4}$s$^{-1}$ (Type B) and $5\times 10^{-3}$s$^{-1}$ (Type A). (Reprinted from \cite{Chihab:1987ez} with permission from Elsevier)}. 
\label{fig:osc}
\end{figure}

This model displays stick-slip behavior as in Ref.~\cite{Braun:2002sf} below a characteristic displacement rate. A major prediction of such ``avalanche oscillator" (as it was labeled) is that the stick-slip transition displacement rate should become comparable to $a/\tau$, where $a$ is the average inter-junction spacing ($\sim 1/\sqrt{\rho}$) and $\tau$ is the junction aging timescale. Assuming that in a dislocation forest ($\rho=10^{16}m^{-2}$), $a=10$ nm and also, $\tau=10^{3}$ s (as in usual elastic contacts), it gives a stick-slip transition displacement rate of $10^{-2}$ nm/s. Furthermore, another prediction would be that stress fluctuations at moderate strain rates (larger than the stick-slip transition rate which should be less than $10^{-5}$ s$^{-1}$) scale inversely proportional to the square root of the number $N$ of aging dislocation junctions $\delta\sigma\sim 1/\sqrt{N}\sim 1/\sqrt{\rho}$ (since $N\sim\rho$). This is a prediction that has been glimpsed in 2D and 3D-DDD simulations~\cite{Papanikolaou:2015wt, csikor2007dislocation}. The verification of models like the one described above or elsewhere~\cite{papanikolaou2012quasi} for a ``dislocation forest" may take place through the observation and confirmation of such stick-slip oscillatory phenomena.  Figure \ref{fig:osc} shows: (A) the differential displacement curves for various displacement rates in uniaxial compression tests of $Ni$ micropillars~\cite{papanikolaou2012quasi}, and (B) stress-time curves for an AlMg alloy at T = 300 K showing the change from type C to type B, and then to type A serrations with increasing strain rate.

Further tracking the competition between abrupt events and slow dislocation aging phenomena in nanocrystals, minimally generalized interface depinning models~\cite{ni2017probing} have been recently utilized towards explaining viscoplastic relaxations of uniaxially compressed pillars  caused by cyclic loading conditions. In this work, oscillatory loads were imposed in the nominal elastic regime of  uniaxially compressed 500 nm-diameter single crystalline Cu pillars at various applied stresses, always above the bulk yield point of $\sim10 $ MPa to investigate the correlations between strain bursts and fatigue loading. The experiment was explained by a mesoscale dislocation plasticity model, which accounts for fast dislocation avalanches  and the slow viscoplastic response. The scaling analysis shows a smooth transition of the system from perfect elasticity to dislocation depinning-driven plasticity that occurs at loads much lower than the nominal yield stress. 

Last but not least, it is worth mentioning that interface depinning models have been heavily used as toy models in various insightful but basic efforts towards exploring the interplay between disorder and plastic deformations in crystals and other systems, without much focus on mechanical properties such as hardening and strength, or basic connections with microstructural and other microscopic details~\cite{Dahmen:2009kl,Friedman:2012oq,uhl2015universal}.

\subsection{Two Dimensional Dislocation Dynamics Results}
Some interesting results of 2D-DDD have been reported in a single slip dislocation system without any obstacles~\cite{Salman:2011pb,Salman:2012dk,miguel2002dislocation}. In this case,  the mean strain rate at a fixed applied stress decays with time in a power-law fashion, $\langle \dot\gamma(t)\rangle \sim t^{-\theta}$, with $\theta=0.65$ \cite{miguel2002dislocation,Ispanovity:2011ly}. Furthermore, the steady-state strain rate displays a power-law, depinning-like behavior $\langle \dot\gamma\rangle \sim (\tau_{\rm ext}-\tau_c)^{1.8}$ \cite{miguel2002dislocation}.  This system also displays non-trivial bursts for $\sigma<\sigma_c$ that have been studied in detail. In order to quantify them, $\sigma_{\rm ext}$ is increased at a slow rate, and the average dislocation velocity $V(t)=1/N\sum_i | \dot x_i (t)|$ is continuously recorded. When $V(t)>V_{\rm th}$, an avalanche propagates, defining avalanche displacements through the slip steps of all dislocations $s_i\delta x_i$, where $s_i$ are the dislocation charges and $\delta x_i$ denotes the displacement at a single timestep. The total amount of slip until $V(t)<V_{\rm th}$, $S=\sum_{t \; for \; V(t)>V_{\rm th}}\sum_i s_i\delta x_i$, gives the ``size" of a strain burst. The probability distribution of $S$ can be drawn by investigating multiple system realizations for bins of the applied stress $\sigma_{\rm ext}$. It has been argued recently that $P(S)\sim S^{-1.0} f(S/S_0)$ with $S_0(\sigma_{\rm ext},N)\sim N^{0.4} \exp(\sigma_{\rm ext}/\sigma_0)$. The fact that the cutoff $S_0$ increases with system size at very small applied stress as $S_0\sim N^{0.4}$ makes the power-law behavior of this system {\it scale-free} in the limit of a large number of dislocations, similar to earthquakes. This particular model behavior ($S_0\sim N^{0.4}$) has not yet been tested in experiments, and it supports the conclusion that the phenomenology of the elastic interface depinning is not applicable. The principal disagreement involves the fact that quantities such as the event-cutoff sizes or durations are principally limited by the system size and do not seem to display any particular scaling with $\sigma -\sigma_c$ (the analog of $F-F_c$), as expected by interface depinning theories \cite{Ispanovity:2011ly}. Overall, this 2D-DDD single slip system model with its specially chosen initial random condition~\cite{groma2003spatial} is clearly simplistic with respect to the experimental reality, since it does not capture multiple slip systems, dislocation sources, dislocation junctions, finite boundaries, and also it is crucially dependent on the particular protocol for choosing random initial configurations. However, it has been highly insightful towards providing explicit  predictions for experimental comparisons, such as {\it e.g.} on the possibility of distinguishing mean-field depinning behaviors~\cite{Dahmen:2009kl} from other possibilities through the possible contrast between integrated and stress-resolved avalanche size distributions and their exponents~\cite{ispanovity2014avalanches}; experimental observations on this front still remain under debate~\cite{zhang2012scale,Tsekenis:2011nx,Friedman:2012oq,lehtinen2016glassy}. Furthermore,  this  2D-DDD single slip system has been insightful towards  cyclic loading studies~\cite{ni2017probing,Laurson:2012pi}

When both dislocation sources and obstacles are included in finite samples \cite{Papanikolaou:2015wt}, however, the material displays phenomena analogous to experimental observations. In this minimal model, yield strength size effects, as well as stochastic effects of post-yield plastic flow are present. The critical ingredients for such predictions are the large density of obstacles, the relative strength of dislocation sources, and the rather small spacing between potentially active slip planes of just 10 b. When sources are much weaker than obstacles, strengthening with decreasing width is consistent with the experimentally observed scaling $\sigma_{Y}\sim w^{-0.45}$. The statistical distribution of events $P(S)$ acquires a power-law tail in the limit of small widths with an exponent $\tau=1.5\pm0.2$ in $P(S)\sim S^{-\tau}$, but when dislocation sources are comparable in strength or stronger than the obstacles, the strength is virtually independent of width or aspect ratio. However, the statistical distribution of plastic events appears universal across width and aspect ratio, scaling with $\tau=1.9\pm0.2$. It is characteristic that this model  demonstrated a clear connection to interface depinning phenomena: at small activation strengths (compared to  obstacle strengths) for dislocation sources, {\it assisted dislocation depinning} takes place at large widths, that is strongly associated with dislocation pile-ups.  Then, the behavior switches to an {\it unassisted dislocation depinning} mechanism at small widths, where dislocations jump over obstacles individually without being assisted by dislocation pile-ups. 

	\begin{figure}[h!]
	\centering
	\includegraphics[width=.75\textwidth]{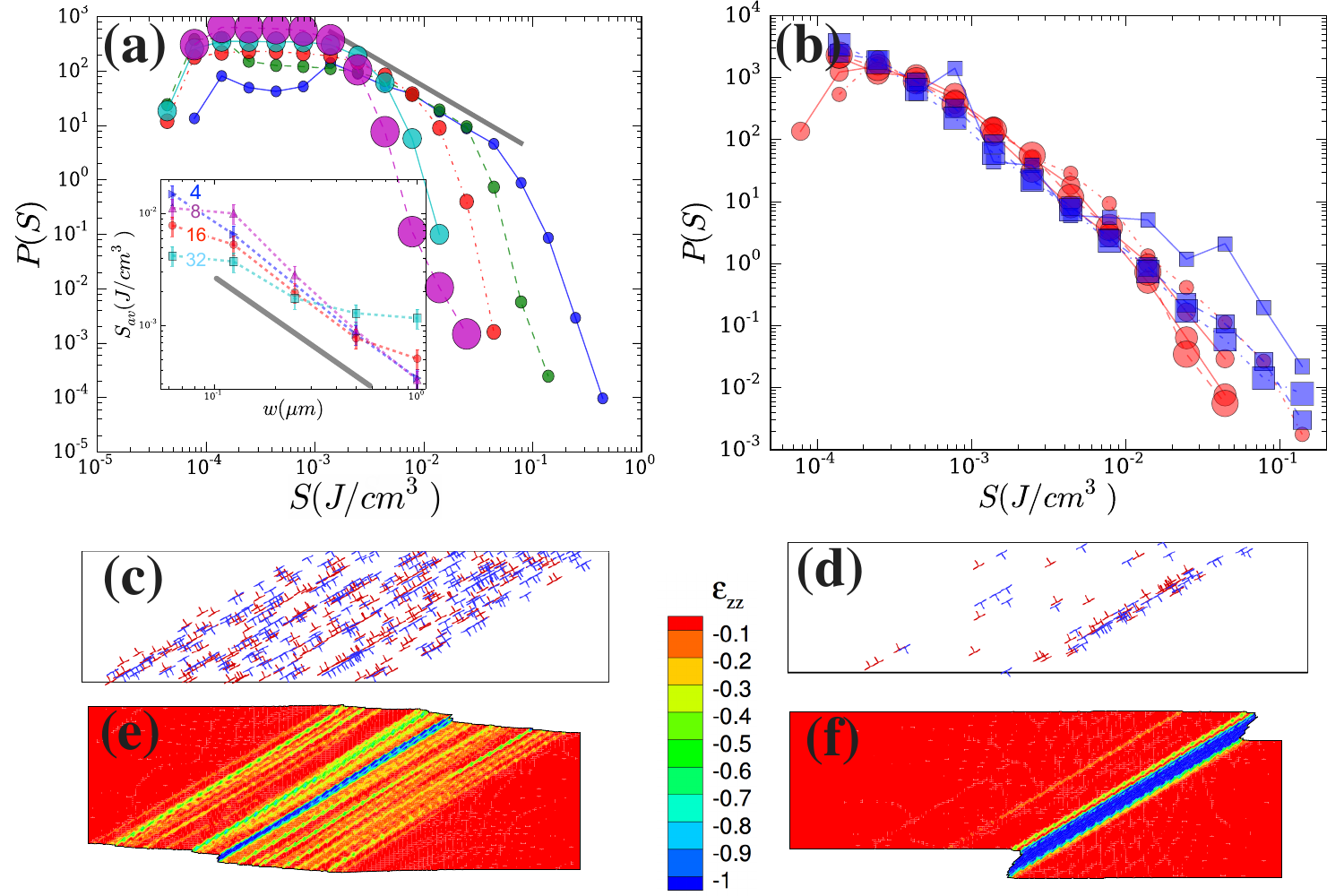}
	\caption{ Uniaxial compression of micropillars using 2D-DDD for relatively weak/strong sources, demonstrating a strong correlation between qualitative features of abrupt event distributions and strain pattern formation~\cite{Papanikolaou:2015wt}. (a,c,e) show abrupt event (stress drops) distributions and patterning for randomly placed FR sources with nucleation stress $50$MPa, while (b,d,f) show the same for FR sources with nucleation stress $300$MPa. Randomly placed obstacles have critical stress $300$MPa.} 
		\label{fig:soc}
	\end{figure}

The crossover that was identified in Ref. \cite{Papanikolaou:2015wt} naturally leads to strengthening, but also it leads to the onset of critical avalanches in the limit of small widths. Stochastic plastic flow fluctuations show a quadratic dependence on the yield stress $\delta\sigma_f\sim \sigma_{Y}^{1.84}$, due to the unassisted dislocation depinning mechanism. It is characteristic that this mechanism clearly distinguishes nanopillar crystals from other materials, which display abrupt plastic flow, such as bulk metallic glasses \cite{zhang2005effect}. Figure \ref{fig:soc} (left column) shows the results of 2D DDD simulations of single-slip for weak sources (50 MPa) in a landscape of strong obstacles (300 MPa). The right column shows results for comparable sources and obstacles (300 MPa).  The figure displays: (a) The width dependence of the event size distribution $P(S)$ (with $S$ being proportional to stress drops in displacement controlled simulations), demonstrating a clear power-law distribution as width decreases for aspect ratio $\alpha=4$ (here, symbol size reflects the width $w$). In the inset, the average event size is shown as a function of $w$ for different aspect ratios $\alpha$. (b) Event statistics for a large nucleation source strength, for various $w$ and $\alpha$. Universal behavior is observed in the event sizes with $P(S)\sim S^{-2.0}$. (c) Discrete dislocation configuration in the weak-source case at 5$\%$ strain. (d) Discrete dislocation configuration in the strong-source case at $5\%$ strain. (e) Strain profile in the weak-source case at $5\%$ strain. (f) Strain profile in the strong-source case at $5\%$ strain. 

\subsection{Results of Reaction-Diffusion and Other Continuum Models}
The main behavior that reaction-diffusion models describe is in the context of the Portevin-Le Chatelier phenomenon. These models have demonstrated that they may predict various dynamical phenomena, such as complex bifurcations, limit cycles, and relaxation oscillations.  One concrete example of the applicability of these dynamical models to the understanding of the PLC phenomenon is the demonstration of the cross-over dynamics from oscillations to chaos by tuning the strain rate.  To understand this, one has to analyze a stress signal time-series under applied strain rate, and determine the underlying dynamics.  Several methods have been reviewed by Ananthakrishna~\cite{Ananthakrishna:2007lq}, however, we will describe the method of Lyapunov spectral analysis and show its application here.

\begin{figure}[htbp]
	\centering
	\includegraphics[width=.6\textwidth]{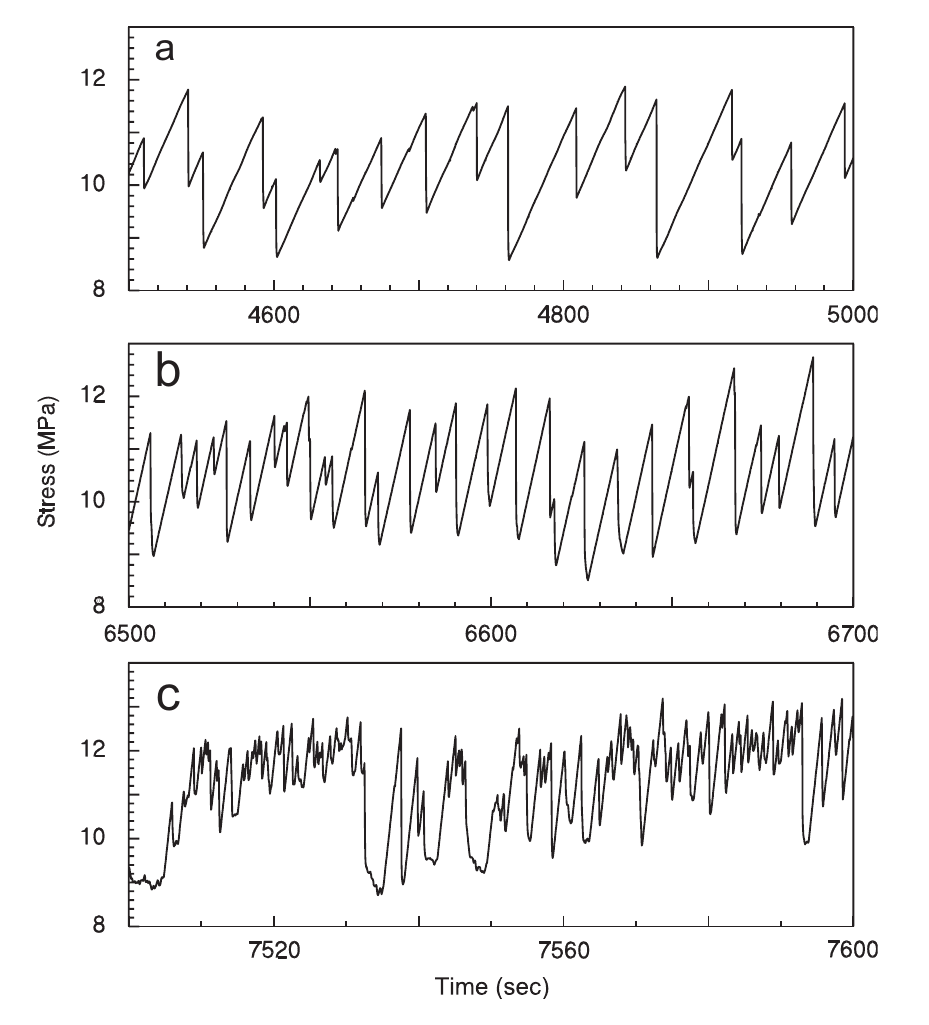}
	\caption{ (a,b,c) Experimental stress-strain curves for $\dot \varepsilon_a=3.3 \times10^{-6} s^{-1}$,  $1.7 \times10^{-5} s^{-1}$  and  $8.3 \times10^{-5} s^{-1}$ (Reprinted from \cite{ananthakrishna1999crossover} with permission from American Physical Society)} 
		\label{fig:Anantha1}
	\end{figure}

Eckmann {\it et al.} \cite{eckmann1986liapunov}  proposed a method of estimation for the Lyapunov spectrum from time series, relying on the construction of a sequence of tangent matrices {\bf T}$_i$, 
which map the difference vector $\vec \zeta(j+k)-\vec\zeta(i+k)= {\bf T}_i (\vec\zeta(i)-\vec\zeta(j))$, where k is the evolution time in units of the time step $\Delta t$. The main idea behind the Eckmann method is to resolve the evolved difference vector  $(\vec\zeta(i)-\vec\zeta(j))$, which aligns itself in the direction of maximum stretching, for a few neighboring points lying within a certain shell in various directions. The procedure determines the elements of the tangent matrix {\bf T}$_i$, where an average is taken over the entire attractor by re-orthogonalizing the tangent matrices using the {\bf QR} decomposition. Here, {\bf Q} is an orthogonal matrix and {\bf R} is an upper triangle matrix with positive elements. Then, the Lyapunov exponents $\lambda_l$ are given by:
\bea
\lambda_l=\frac{1}{k\Delta t p}\sum\limits_{j=0}^{p-1}\ln({\bf R}_j)_{ll},\quad l=1,2,...,d,
\eea
where p is the number of available matrices and $k\Delta t$ is the propagation time, and d the number of Lyapunov exponents. To examine the possibility of a crossover from chaotic to power law state of stress drops requires the data to be obtained over a wide range of strain rates. In single crystals, the data sets were obtained from Cu-10\%Al \cite{ananthakrishna1999crossover} samples. The crystals were oriented for easy glide. The deformation tests were carried out at 620K under three different strain rates $3.3 \times10^{-6} s^{-1}$,  $1.7 \times10^{-5} s^{-1}$  and  $8.3 \times10^{-5} s^{-1}$. Figure \ref{fig:Anantha1} show the stress drop time series for  PLC. The Lyapunov spectrum calculated by Ananthakrishna \cite{ananthakrishna2007current} using the Eckmann's algorithm indicated that the time series of the low and medium strain rate PLC time series are of chaotic origin \cite{ananthakrishna1999crossover}.
%(l, m and h for ``low", ``medium" and ``high" strain rates)

In addition to well studied reaction-diffusion models, there have been several recent continuum modeling efforts that concentrate on the observations of plastic event probability distributions and the possibility of identifying  a ``universality class" and the related interface depinning exponents, such as $\tau$ and $\alpha$~\cite{Koslowski:2007la,Koslowski:2004fy,Chan:2010tg,Fressengeas:2009jt}, but without any grasp and understanding on the distributions' dependence on various model parameters such as the yield strength and hardening coefficient. Due to the extreme timescale separation between loading and abrupt events, these models are plagued either by ultra-simplified dynamics or extreme loading rates that should not relate to experimental evidence. 

\subsection{Results of 3D-DDD Simulations}

3D-DDD simulations are much more demanding computationally than any other method. However, they have recently been robust enough to be utilized as virtual experimental tools for studies of strain bursts and dislocation avalanches.  Several studies have focused on the differences between micro-pillar compression/ tension and bending, on the role of load versus displacement mode control, on the type of statistics (e.g. Weibull versus power law), and on the influence of crystal structure (FCC versus BCC).

Csikor {\it et al.} \cite{csikor2007dislocation} carried out 3D-DDD simulations on Al with characteristic length ranging from 0.5 $\mu$m to 1.5 $\mu$m. The avalanche strain was found to follow power law distribution with an exponent about -1.5; in the range of experimental results \cite{dimiduk2006scale}. A high-density of indestructible Frank-Read sources was used as the initial configuration. Most of the plastic deformation was found to occur on one of four equivalent slip systems, and the spatial distribution of avalanches was found to be lamellar in shape. To study the effects of stress state, Motz {\it et al.}~\cite{motz2008micro,zaiser2013statistical} investigated the strain burst behavior in micro-sized bending beams, and found that each burst corresponds to a rapid increase of the dislocation density. However, no significant dislocation density change was observed under uniaxial compression tests. This suggests that the increase in dislocation density during each burst in bending tests is mainly associated with the plastic strain gradient, which accumulates geometrically necessary dislocations. It is found that irrespective of the existence of strain gradients and the increase in dislocation density, the strain increment for both bending and compression tests  follow the same power law distribution. These results illustrate the robustness of the power law scaling irrespective of external loading. 

\begin{figure}[h!]
		\centering
	\includegraphics[width=.7\textwidth]{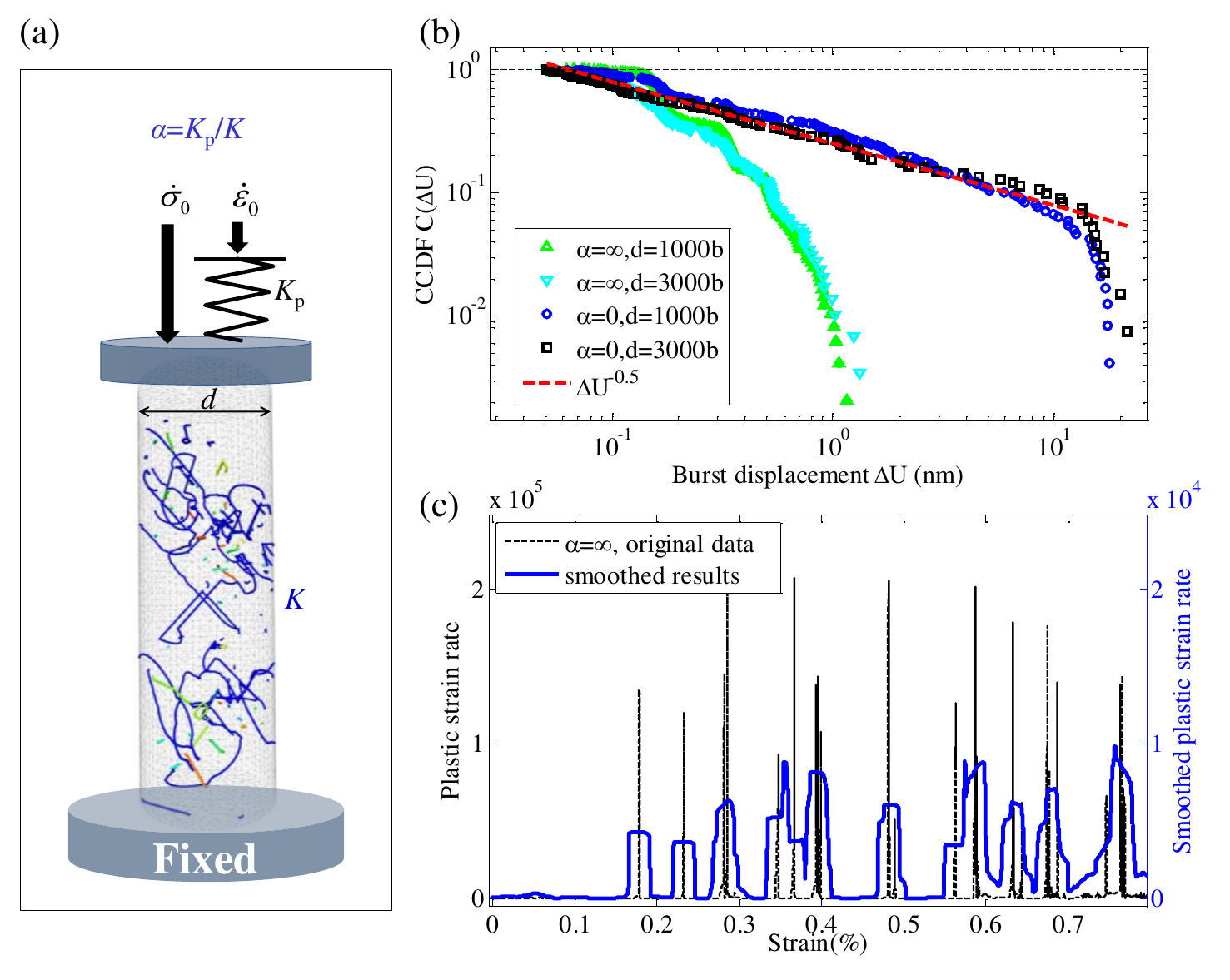}
	\caption{(a) Simplified sketch of pillar compression with an open-loop (directly applying a stress rate $\dot{\sigma}_0$) and closed-loop control (connecting to a spring with a finite machine stiffness $K_p$ to realize displacement control with strain rate $\dot{\varepsilon}_0$). $\alpha$ is a dimensionless stiffness ratio between $K_p$ and sample stiffness $K$. (b) Complementary cumulative distribution function of burst displacement under pure strain control ($\alpha=+ \infty$) and pure stress control ($\alpha=0$) for pillars of different diameter $d$. (c) Typical results of the evolution of plastic strain rate and its averaged value in 0.24 $\mu$s windows, showing quasi-periodic strain bursts under pure strain control (Reprinted from \cite{ cui2016avalanche} with permission from American Physical Society)}
		\label{fig:yinan}
	\end{figure}

The influence of the external system loading mode on strain bursts and dislocation avalanches has been the subject of recent investigations \cite{cui2016avalanche,cui2016Influence}. It was found that the loading mode can induce a controllable dynamical regime transition from SOC avalanche power-law scaling (see Fig. \ref{fig:yinan}b) to quasi-periodic strain burst oscillations (see Fig. \ref{fig:yinan}c). The differences in correlated dislocation activities in both cases, related to slow or fast stress relaxation, were revealed. These results point out to new possibilities for novel experiments with a faster response rate than currently obtainable, and which can be designed to explore this dynamical transition regime.

3D-DDD simulations have also been used to understand the statistical distributions of strain bursts~\cite{Devincre:2008zm}. In simulations of submicron FCC crystal plasticity, it was found that the weakest source mechanism is dominant \cite{el2009role,senger2011aspect}. Strain bursts were found to be dominated by the intermittent operation of the weakest sources  \cite{rao2008athermal,cui2014theoretical}. However, power law statistics were found to be the result from correlated activation of dislocation sources, and that such correlations are strongly influenced by the loading mode \cite{cui2016Influence,crosby2015origin}. 

The effects of dislocation barriers on strain bursts have been investigated, mainly in the context of radiation-induced barriers, surface coatings, and grain boundaries.  If irradiation defects are considered, their resistant stress was found to inhibit strain bursts. However, the gradual destruction of irradiation defects may promote larger bursts \cite{Cui2016irradiation}. In addition, the effects of interfaces, such as coatings \cite{el2011trapping,cui2015theoretical} and grain boundaries \cite{huang2017extended} on burst behaviour were also considered. Devincre {\it et al.} \cite{devincre2008dislocation} studied the dislocation avalanche behavior in FCC Cu crystals with cell size about 5 $\mu$m and periodic boundary conditions. It was found that the formation and destruction of junctions between glide and forest dislocations control dislocation avalanches in bulk crystals. 

3D-DDD simulations with periodic boundary conditions were carried out on FCC Al crystals of size in the range 0.7- 2.15 $\mu$m \cite{lehtinen2016glassy}, revealing a power law scaling avalanche behavior.  From these observations, it was proposed that strain bursts revealed by 3D simulations do not follow depinning-like non-equilibrium phase transitions, but display an extended, critical-like phase. It is more possible to observe a depinning-like scenario in obstacle dominated plasticity, and that may explain the conclusions of mostly 2D-DDD simulations with obstacle strengthening. In crystals with randomly distributed immobile solute atoms, 2D-DDD results demonstrated that when the disorder strength is low, a power law distribution is found for both avalanche size and duration. However, for very strong disorder, critical dynamics creases due to a strong pinning-induced avalanche size-limiting effect \cite{ovaska2015quenched}, consistent with experimental results \cite{zhanga2016taming}. 

\section{The Plasticity-Fracture Connection}
\label{sec:pf}
	
Connecting plasticity with fracture constitutes a long-standing challenge for material science and fracture mechanics, especially due to the intrinsic multi-scale nature of the fracture process. Abrupt deformation processes in nature are typically connected with catastrophic events. Earthquakes \cite{xia2004laboratory} is the most common example, where abrupt fault dynamics and fast sliding leads to further fragmentation and fracture, with associated well known scaling laws such as the Gutenberg-Richter law \cite{gutenberg2013seismicity}. In crystal deformation, the analogous exponents associated with acoustic emission events corresponding to microcrack formation are generally around $1.3 \sim 2$ for wood, fiberglass, paper, polyurethane foams {\it et al.} \cite{deschanel2009experimental}. In addition, scaling invariance is also observed for temporal (waiting time, calming time) and spatial (fractal structure of the rupture) distributions \cite{bak2002unified,christensen2002unified}. The extensive experimental work about the fractal self-affinity and scaling property of fracture surfaces in metal and alloys are reviewed in detail by Bouchaud \cite{bouchaud1997scaling}.

The influence of plasticity on fracture statistics is seldomly studied. It is found that the acoustic emission signals produced by dislocation-governed deformation and intercrystalline fracture are distinctly different \cite{baram1981effect}. This leads to the possibility to differentiate the signals by the acoustic emission technique, and then simultaneously investigate the statistics of cracks and plasticity and their interactions. Various efforts have been made to formulate a statistical thermodynamics of fracture and build the simplified models to understand its underlying statistics \cite{zapperi1999avalanches,herrmann2014statistical}. Block-spring arrays (stick-slip model) and cellular automata models are mostly used to simulate the collective behavior of earthquack and faults \cite{burridge1967model,otsuka1972chain,wolfram1984cellular}. Based on an analytical failure model of bundles of parallel fibers, the criticality character of fracture in brittle material is captured as it approaches breakdown \cite{hansen1994criticality}. In addition, large-scale simulations of lattice models are widely used to understand the statistics of fracture in disordered system \cite{herrmann2014statistical}, such as the acoustic emission during the opening of crack in hydraulic fracturing \cite{tzschichholz1995simulations}, and the acoustic energy bursts produced by microcrack during dynamic fracture \cite{minozzi2003dynamic}. Power-law scaling of avalanche distribution were observed, consistent with experimental observations \cite{cannelli1993self}. Based on mean-field calculations and numerical simulations, Zapperi {\it et al.} revealed that the breakdown in disordered media can be described by a first-order phase transition, similarly to thermally-activated homogeneous fracture \cite{zapperi1999avalanches}. They further stated that SOC assumes a slowly driven system with a critical stationary
state, while the fracture problem has no stationary state \cite{zapperi1997first}.  Thus, power-law scaling is not self-organized because the control parameter is externally ''swept" towards the instability \cite{sornette1994sweeping}.

Two main questions must be answered to reveal the connection between localized plastic deformation and fracture. The first one is: how does plastic deformation affect the initiation of cracks? Roughly speaking, crack initiation requires that the local normal stress of the crack plane is higher than a specific value, and that crack formation is accompanied by a reduction of the system energy \cite{stroh1954formation}. Cracks generally prefer to initiate at a stress concentration region, for example induced by the highly localized plastic deformation, or pile-up groups of dislocations. Therefore, crack initiation is sensitive to the spatial distribution of plastic deformation, and to dislocation patterns. The other question is: what is the role of plasticity for crack growth? More efforts have been undertaken to investigate this problem, due to its significance in improving material ductility.~It is generally believed that the nucleation and movement of dislocations near the crack tip will blunt the crack and inhibit further crack propagation. Considering that plasticity behaves as a ''crackling noise" with intermittent bursts and dislocation avalanches characterized by scale-free size distributions and self-organized pattern formation, it will be interesting to check the dislocation avalanche behavior under the high concentrated stress field induced by a crack, and its feedback on crack propagation.

While the abruptness of the deformation originates in its effectively athermal nature, the ``damaging" character clearly originates in the ``size" of the deformation which is directly connected to the impact strength.  Naturally, therefore, in crystal plasticity, it is expected that the abruptness of deformation may be directly connected to critically important phenomena such as strain localization, dislocation patterning and stick-slip instabilities. Is it possible to predict such phenomena by estimating the character of plastic fluctuations? While this subject lies along the research frontiers, it is evident that such possibilities may exist.  The simplest  phenomenon to understand is the relationship between an edge dislocation pile-up at a single obstacle and fracture initiation.~After crossing the obstacle, dislocations in the pileup  leave the crystal and deposit discrete steps at nearby surfaces or grain boundary interfaces. Dislocation structures evolve into spatial patterns, such as cell walls~\cite{Kawasaki:1980fk}, and labyrinth or vein structures~\cite{Mughrabi:1983uq}. The characteristic length scale of such patterns is typically proportional to the dislocation spacing $\sim 1/\sqrt{\rho}$ and inversely proportional to the stress; the so-called similitude relationships~\cite{Sauzay:2011kx}.  The precise way in which such pattern formation is connected to the abrupt character of the dynamics still needs further investigation. 
	
In irradiated crystals, it has been clear that glide dislocations strongly interact with irradiation defects, gradually leading to the formation of defect clear channels. This quickly serves as patterns of localized deformation zones, where most of the macroscopic plastic strain is distributed, while vast regions of the material have hardly any strain. Such kind of plastic instability due to spatial patterning generally occurs in high-dose irradiated crystals \cite{ghoniem2001dislocation,victoria2000microstructure,byun2004temperature}. The mechanisms of dislocation channel formation have been investigated by 3D-DDD simulations \cite{de2000multiscale,arsenlis2012dislocation,Cui2017irrachannel}.  Although the exact connection between localized plastic deformation in irradiated materials and their easy fracture has not yet been clarified, the conditions for the emergence of localized plastic flow in small size samples has recently been delineated \cite{Cui2017irrachannel}.

	The basic observation that makes such connections rather plausible in crystal plasticity is the fact that fractal slip surface steps have been typically observed in plastically deforming crystals \cite{neuhauser1988dynamics,Hahner:1998uq}. Fractals demonstrate a correlation of the microstructural response and possibly avalanche phenomena. However, it is the similitude principle \cite{kubin1993dislocation} that connects spatial and temporal behavior (in this way, the spatial power law fractality naturally implies a temporal power law fractality as well). In the case of such a behavior, it is natural to expect that plastic flow localization effects and other fracture-inducing phenomena should alter the probability distribution for the experimentally observed avalanche burst phenomena. 
	
\section{Conclusions and Future Prospects}	
In our review of the plethora of experimental observations, theoretical models, and computational simulations related to the subject of strain bursts and dislocation avalanches, we emphasized the universality of fundamental phenomena underlying plasticity and fracture.  It is generally the realm of hard engineering to describe, at a continuum level, the impact of plastic deformation and fracture on design of engineering components. Nevertheless, it is abundantly clear that these phenomena, when viewed from a statistical mechanics lens, offer a rich variety of physical behavior that is deeply rooted in the physics of non-linear dynamics, phase transitions, critical systems, and universality classes.  This makes it the more intriguing, as we, on the one hand, attempt to link plasticity and fracture to general physical phenomena; yet, on the other hand, try to utilize such physics in designing better materials.

The vast experimental observations reviewed here show that strain bursts and dislocation avalanches are very prevalent in diverse systems. Their influence is manifest in small size materials at the nano-scale, as well as at the macro-scale, where dislocation avalanches turn destructively into massive localized plastic deformation that leads to fracture. At the nano-scale, it turned out that there is some uniqueness that is particular to the small size, as evidenced by experiments on nano-indentation and other experiments on micro-pillars.  The results of any experiment at this small scale are not consistently repeatable, but they nevertheless obey the laws of statistics. Our experience with macroscopic experiments on the mechanical deformation of materials is such that the strength, or hardening characteristics, can be determined as fixed quantities, usually associated with experimental error that can be quantified. At the nano- and micro-scales, however, we see from the review here that such a description is totally insufficient, and that the mechanical properties are best described as statistical distribution functions.

Statistical analysis of nano- and micro-system behavior, revealed by nano-indentation and micropillar compression tests, have revealed a consistent picture of strain bursts and dislocation avalanche physics.  In pristine or ``starved" nano- and micro-systems, the release of elastic energy in strain bursts is governed by weak-link statistics.  Because of the scarcity of dislocation sources, they are activated almost independently of one another, and dislocation segments of the longest length are inevitably activated when a stress is applied.  Once activated, a few correlated events may take place, but generally, Weibull statistics will govern the distribution of weak links, in a similar fashion to brittle fracture. On the other hand, in heavily-deformed small crystals, a great degree of cooperation between activated dislocations is evident from the nature of statistics.  Under stress rate control, dislocation sources are activated sequentially as the stress rises, and a domino effect takes place, where groups of dislocations move simultaneously in a true avalanche behavior.  The statistics governing this system dynamics is of the ``scale-free'' power-law type, with exponents around $\sim1.5$. However, even though this scenario is well-supported by the vast majority of experimental data and theoretical models, intricate features still emerge:  First, it appears that the external loading mode may have the ability to \emph{tune} the system dynamics. When the loading mode is controlled by the stress rate, dislocation avalanches are continuously activated giving rise to scale-free statistics, and the usual effects of system size on the tail of the power law distribution.  On the other hand, displacement rate control, which is supposed to tune a depinning system towards criticality, appears to display that every release of strain energy  to immediate stress relaxation, effectively shutting down the operation of critical sources. The dynamics of the system appears to change qualitatively from scale-free to relaxation-oscillations.

It is very interesting to note that dislocation ensembles, in small and in large volumes, can provide a vivid laboratory for studies of non-linear dynamics and cooperative phenomena.  The tunability of system dynamics, demonstrated in micropillars, via the external load control is one example of dynamical transitions that can directly be measured experimentally. Likewise, in large sample volumes in alloys that exhibit the PLC effect, the strain rate itself has been conclusively shown to tune the qualitative dynamics from limit cycles to chaotic behavior. In this regard, one can indeed see that the temporal dynamical characteristics associated with strain bursts and dislocation avalanches are inextricably coupled with spatial patterning and self-organization that is often observed in plastically-deformed materials. A stark example of this coupling of space-time is the observation of intense deformation channels in irradiated materials associated with clearing of radiation-induced defects. Similar observations have been reported in precipitation-hardened alloys.  The common thread linking these phenomena is the limited stability of dislocation barriers, where under irradiation, small interstitial loops, nano-voids, or stacking fault tetrahedra, can be destroyed by glide dislocations, in much the same way as shearable precipitates in hardened alloys.  This commonality leads to spatial defect or precipitate-free patterns that mirror the temporal structure of stress relaxation signals under strain control.

It may soon be within our reach that we could build materials by design that control the homogeneity of strain bursts and dislocation avalanches. For example, for fiber-structured materials, one may imagine that avalanches get ``arrested". In a similar way, it could happen that similar ``arrest" takes place in nano-grained materials at grain boundaries.  It is therefore tantalizing to think of how we can exploit the physics of strain bursts and dislocation avalanches to design the ultimate materials that are extremely strong and ductile at the same time!  The avenues to achieve such possibilities would have to involve new methods of fabrication that can produce uniform and stable distributions of ``arrest centers," such as nano-precipitates, engineered grain boundaries around nano-grained materials, laminated structures (e.g. lath boundaries, twin boundaries, interfaces introduced by severe plastic deformation, etc.).  While these speculations depend on concrete advances in manufacturing and material processing technologies, there are a number of remaining challenges that await further experimental, theoretical, and modeling investigations. Some of such challenges are:

	\begin{enumerate}
	\item The initial dislocation configuration influences the character of observed phenomena, as in any system far from equilibrium. This would seem to emphasize the probabilistic and statistical aspects of strain bursts and dislocation avalanches that cannot be ignored.
	\item The dependence of avalanche phenomena on system size is still not fully explored. Some progress has been made in small systems (e.g. nano- and micro-pillars, nano-indentation). However, the gap between this length scale and macroscopic dimensions is huge, presenting a chalenge to the modeling community.
	\item The fractal step distribution has been explored in bulk materials, yet it still remains a question on how it translates at the nanoscale. Is there a critical length scale and timescale where such fractals are not visible anymore? Is such a length scale related to fracture and crack initiation?
	\item Aging phenomena, either in the context of dislocation dynamics or precipitate/solute-related dynamic strain aging and other ``slow" strengthening and weakening phenomena must be accounted for in microscopically-informed simulations such as DDD. This presents a modeling challenge to 3D-DDD.
	\item The inclusion of dislocation climb processes due to point defect flux in 3D-DDD is still an open question.  The complication is that the timescales of dislocation glide versus climb are vastly different.
	\item Finally, the appropriate description of the complexity of the mechanisms involved in strain bursts and dislocation avalanches remains as an an open question.  Direct numerical simulations try to deal with physical complexities at the expense of clear and simple descriptions, while conceptual toy models give a simple and clear description of the physics at the expense of reality.  Striking a balance between these complementary but opposing approaches may just be the best approach!

	\end{enumerate}

%%%%%%%%%%%%%%%%%%%%%%%%%%%%%%%%%%%%%%%%%%%%
\section*{Acknowledgements}

This work is partially supported by the Air Force Office of Scientific Research (AFOSR), Award No: FA9550-16-1-0444 with UCLA, by the US Department of Energy, Award Numbers  DE-FG02-03ER54708(NG and YC) and  DE-SC0014109 (SP), and by the Department of Commerce - NIST, Award Number 1007294R (SP).
%%%%%%%%%%%%%%%%%%%%%%%%%%%%%%%%%%%%%%%%%%%
\newpage
\bibliographystyle{unsrt}
%\bibliography{./Biblio/references_burst,./Biblio/references_burst_SP,./Biblio/Part2-Refs}

%%%%%%%%%%%%%%%%%%%%%%%%%%%%%%%%%%%%%%%%%%%

\end{document}